\documentclass[traditabstract]{aa}
\usepackage{txfonts}

\usepackage{color}

\usepackage{graphicx}

\usepackage{rotating}
\usepackage{amsfonts}

\usepackage{pdflscape}

\usepackage{subfig}

\usepackage{natbib}
\bibpunct{(}{)}{;}{a}{}{,} 

\begin{document}

\title{Long-term magnetic activity of a sample of M-dwarf stars from the HARPS program\thanks{Based on observations made with the HARPS instrument on the ESO 3.6-m telescope at La Silla Observatory under programme ID 072.C-0488(E)}}
\subtitle{II. Activity and radial velocity}

\author{J. Gomes da Silva\inst{1,2}
	\and N.C. Santos\inst{1,2}
	\and X. Bonfils\inst{3}
	\and X. Delfosse\inst{3}
	\and T. Forveille\inst{3}
	\and S. Udry\inst{4}
	\and X. Dumusque\inst{1,4}
	\and C. Lovis\inst{4}
	}

\institute{Centro de Astrof\'isica, Universidade do Porto, Rua das Estrelas, 4150-762 Porto, Portugal \\ \email{Joao.Silva@astro.up.pt}
	\and Departamento de F\'isica e Astronomia, Faculdade de Ci\^encias, Universidade do Porto, Portugal
	\and UJF-Grenoble 1 / CNRS-INSU, Institut de Plan\'etologie et d'Astrophysique de Grenoble (IPAG) UMR 5274, Grenoble, F-38041, France
	\and Observatoire de Gen\`eve, Universit\'e de Gen\`eve, 51 ch. des Maillettes, CH-1290 Versoix, Switzerland
	}

\abstract{
Due to their low mass and luminosity, M dwarfs are ideal targets if one hopes to find low-mass planets similar to Earth by using the radial velocity (RV) method.
However, stellar magnetic cycles could add noise or even mimic the RV signal of a long-period companion.
Following our previous work that studied the correlation between activity cycles and long-term RV variations for K dwarfs we now expand that research to the lower-end of the main sequence.
Our objective is to detect any correlations between long-term activity variations and the observed RV of a sample of M dwarfs.
We used a sample of 27 M-dwarfs with a median observational timespan of 5.9 years.
The cross-correlation function (CCF) with its parameters RV, bisector inverse slope (BIS), full-width-at-half-maximum (FWHM) and contrast have been computed from the HARPS spectrum. The activity index have been derived using the \ion{Na}{i} D doublet.
These parameters were compared with the activity level of the stars to search for correlations.
We detected RV variations up to $\sim$5 m\,s$^{-1}$ that we can attribute to activity cycle effects.
However, only 36\% of the stars with long-term activity variability appear to have their RV affected by magnetic cycles, on the typical timescale of $\sim$6 years.
Therefore, we suggest a careful analysis of activity data when searching for extrasolar planets using long-timespan RV data.
}



\maketitle

\section{Introduction}
The majority of extrasolar planets discovered so far were either detected or confirmed via the radial velocity method\footnote{cf. http://exoplanet.eu/}.
This technique measures the Doppler effect due to the wobble of the star around the centre of mass of the star-planet system.
As an indirect method, it is sensitive to stellar sources of noise like oscillations, granulation, rotating active regions, and magnetic cycles.

Stellar oscillations and granulation induce radial velocity variations on timescales up to some hours and can easily be suppressed if a optimized observational strategy is used \citep{dumusque2011b}.
However, this is not true for the case of rotationally modulated active regions which have longer timescales.
The effect of rotating active regions on line profiles, and hence RV measurements, can easily hide or mimic the signal of orbiting companions \citep[e.g.][]{saar1997a,santos2000,queloz2001}.
In some cases, these effects can be diagnosed e.g. by the anti-correlation between instantaneous measurements of RV and the bisector inverse slope \citep{queloz2001,boisse2011} and corrected by subtracting the anti-correlation slope to the RV measurements \citep[e.g.][]{melo2007,boisse2009} or by fitting an extra Keplerian orbit to the RV data with the period detected in the activity time-series \citep[e.g.][]{bonfils2007,forveille2009}.
\citet{queloz2009} and \citet{boisse2011} proposed to fit the activity signal with sinusoids of different periods: the rotation one and its harmonics.
\citet{saar2000} used a different technique to correct for long-term activity induced RV. These authors used the slope of the $S_{IR}$--RV correlation to remove the activity influence from the radial velocity signal.

Long-term stellar magnetic cycles can also be a source of noise for precise RV measurements.
\citet{kurster2003} studied the correlation between radial velocity and the H$\alpha$ index for the nearby Barnard's star (Gl\,699).
They found an anti-correlation between velocity and the activity index with a correlation coefficient of $\rho = -0.50$. The authors concluded that activity as measured by the hydrogen line was producing a blueshift of the photospheric absorption lines.

By using simulations, \citet{meunier2010} showed that magnetic activity cycles can induce radial-velocity variations for the case of the Sun as seen edge-on with amplitudes that can reach the $\sim$10 m\,s$^{-1}$ level.
These kind of variations in a star with a periodic activity cycle could be able to mimic the signal of a long-period extra-solar planet.

In this context, \citet{santos2010} used a sample of 8 solar-type stars with the aim of studying if a correlation between long term activity and RV variations exists. The long-term variations were detected on the $S_{MW}$, H$\alpha$, and \ion{He}{i} indices and the BIS, FWHM and contrast CCF parameters but only two stars showed correlations with RV stronger than $\rho > |0.75|$: one positive correlation and  one anti-correlation. The authors concluded then that the possible amplitudes of induced RV variations for the early-K dwarfs was low, at the $\sim$1 m\,s$^{-1}$ level similar to the HARPS precision.

Using a larger sample of around 300 stars from the HARPS FGK high precision program, \citet{lovis2011} studied the correlation between long-term activity variations with radial-velocity in a similar fashion as \citet{santos2010} did.
They found a correlation between the slope of the RV--activity index ($R'_{HK}$) correlation, effective temperature ($T_{\mathrm{eff}}$), and metallicity ([Fe/H]).
The slope is weaker for late-type dwarfs than for early ones and therefore, it seems that the RVs of later-K dwarfs are less affected by magnetic cycles than the RVs of early-G dwarfs.
Therefore, the influence of long-term activity could be corrected if the activity level, effective temperature and metallicity of the star are known by using the slope of the RV--$\log(R'_{HK})$ relation.

Hints of long term RV variations produced by activity cycles were found by \citet{moutou2011} in the stars BD-114672 and HIP21934, with periods of 1692 and 1100 days respectively. \citet{segransan2011} also found a long term period of $\sim$500 days on the radial velocity of HD104067 (K2V). The activity index of this star, hosting a 55-day period Neptune-like planet, was found to have a correlation with the RV residuals (with $\sigma_{RV} = 4.6$ m\,s$^{-1}$) after the planetary signal was removed.

More recently, three exoplanets were discovered in three early-K stars with magnetic cycles by correcting the activity signals in the radial velocity data \citep{dumusque2011a}.
The planets were found by fitting simultaneously two keplerians: one for the planet and one for the magnetic cycle. All the parameters of the keplerian fitting the cycle, except the amplitude, were fixed to the ones obtained when fitting the activity index only.
This proves that (i) magnetic activity cycles of stars can influence radial velocity \textit{and} hide the signal of long-period planets, and (ii) a correction of the long-term activity-induced RV is possible and can be used to recover the embedded signal of a planet.

In \citet[][hereafter Paper I]{gomesdasilva2011} we compared the long-term activity variations using four activity indices for a sample of M-dwarf stars from the HARPS program.
We arrived to the conclusion that the \ion{Na}{i} index was the most appropriate of the four to study the activity of these type of stars.
We will now use in this paper the \ion{Na}{i} index as the proxy of activity with the aim of comparing long-term activity to RV and some other relevant CCF parameters using the same sample of stars.
This paper will therefore extend the first study in \citet{santos2010} to the case of early-M dwarfs.

This paper is organised as follows: in Sect. \ref{sect:sample} we present our sample and observation log and explain our data analysis, in Sect \ref{vartests} a selection of a subsample of stars with long-term activity variability is done, in Sect. \ref{cycles_fit} we determine which stars have hints of periodic magnetic cycles, in Sect. \ref{rv_s_ccf} we compare long-term activity with radial velocity and the CCF parameters, in Sect. \ref{indstars} we examine individual stars with strong activity-RV correlation and other interesting cases, and finally in Sect. \ref{conclusions} we draw our conclusions from the present work.

\section{Sample and observations} \label{sect:sample}

\begin{table*}[ht]
\caption{Basic parameters and observation log of the sample.}
\label{sample}
\centering
\begin{tabular}{l c l c c c c c c c c c c c c c c c c} \\
\hline
\hline
\multicolumn{1}{l}{Star} &
\multicolumn{1}{c}{Other names} &
\multicolumn{1}{c}{Sp. Type\tablefootmark{a}} &
\multicolumn{1}{c}{$V$\tablefootmark{d}} &
\multicolumn{1}{c}{$V-I$\tablefootmark{e}} &
\multicolumn{1}{c}{$BJD_{start} - 2400000$} &
\multicolumn{1}{c}{\# nights} &
\multicolumn{1}{c}{$T_{span}$} &
\multicolumn{1}{c}{$<S/N>$} &
\multicolumn{1}{c}{$<\sigma_i(V_r)>$} \\
\multicolumn{1}{c}{} &
\multicolumn{1}{c}{} &
\multicolumn{1}{c}{} &
\multicolumn{1}{c}{[mag]} &
\multicolumn{1}{c}{[mag]} &
\multicolumn{1}{c}{[days]} &
\multicolumn{1}{c}{} &
\multicolumn{1}{c}{[days]} &
\multicolumn{1}{c}{(order 56)} &
\multicolumn{1}{c}{[m\,s$^{-1}$]} \\
\hline					
\object{Gl\,1}							&				&	M3V					&	8.57		&	2.13	&	52985.60	&	45	&	2063	&	95.6		&	0.7 \\
\object{Gl\,176}\tablefootmark{\dagger}		&				&	M2.5V				&	9.97		&	2.25	&	52986.71	&	71	&	2146	&	47.6		&	1.2 \\
\object{Gl\,205}						&				&	M1.5V				&	7.92		&	2.08	&	52986.73	&	75	&	1400	&	108.5	&	0.8 \\
\object{Gl\,273}						&				&	M3.5V				&	9.89		&	2.71	&	52986.77	&	62	&	2140	&	57.3		&	0.8 \\
\object{Gl\,382}						&				&	M2V					&	9.26		&	2.18	&	52986.84	&	30	&	1188	&	66.7		&	1.1 \\
\object{Gl\,393}						&				&	M2V					&	9.76		&	2.26	&	52986.86	&	29	&	1897	&	67.0		&	1.0 \\
\object{Gl\,433}\tablefootmark{\dagger}		&				&	M2V					&	9.79		&	2.15	&	52989.84	&	61	&	2245	&	56.6		&	1.2 \\
\object{Gl\,436}\tablefootmark{\dagger}		&				&	M2.5V\tablefootmark{b}	&	10.68	&	2.02	&	53760.83	&	115	&	1532	&	32.9		&	1.4 \\
\object{Gl\,479}						&				&	M3V					&	10.64	&	1.90	&	53158.55	&	58	&	1413	&	43.3		&	1.2 \\
\object{Gl\,526}						&				&	M1.5V				&	8.46		&	2.07	&	53158.60	&	34	&	1843	&	101.6	&	0.8 \\
\object{Gl\,551}						&	Prox Cent		&	M5.5V				&	11.05	&	3.62	&	53152.60	&	42	&	2140	&	21.1		&	1.3 \\
\object{Gl\,581}\tablefootmark{\dagger}		&				&	M2.5V				&	10.57	&	2.53	&	53152.71	&	194	&	2306	&	36.4		&	1.2 \\
\object{Gl\,588}						&				&	M2.5V				&	9.31		&	2.40	&	53152.75	&	32	&	2200	&	75.4		&	0.8 \\
\object{Gl\,667}\,C\tablefootmark{\dagger}	&				&	M2V					&	10.22	&	2.08	&	53158.76	&	173	&	2201	&	42.9		&	1.3 \\
\object{Gl\,674}\tablefootmark{\dagger}		&				&	M3V					&	9.36		&	2.40	&	53158.75	&	44	&	1574	&	70.6		&	0.8 \\
\object{Gl\,680}						&				&	M1.5V				&	10.14	&	2.27	&	53159.71	&	35	&	2269	&	41.2		&	1.4 \\
\object{Gl\,699}						&	Barnard's star	&	M4V					&	9.54		&	2.52	&	54194.89	&	32	&	1291	&	34.3		&	0.8 \\
\object{Gl\,832}\tablefootmark{\dagger}		&				&	M1V					&	8.67		&	2.18	&	52985.52	&	59	&	2424	&	88.9		&	0.8 \\
\object{Gl\,849}\tablefootmark{\dagger}		&				&	M3V					&	10.42	&	2.50	&	52990.54	&	49	&	2423	&	41.9		&	1.3 \\
\object{Gl\,876}\tablefootmark{\dagger}		&				&	M3.5V				&	10.17	&	2.77	&	52987.58	&	67	&	2508	&	42.6		&	1.1 \\
\object{Gl\,877}						&				&	M2.5V				&	10.55	&	2.43	&	52857.83	&	40	&	2636	&	36.5		&	1.2 \\
\object{Gl\,887}						&				&	M2V					&	7.34		&	2.02	&	52985.57	&	63	&	1407	&	131.3	&	0.8 \\
\object{Gl\,908}						&				&	M1V					&	8.98		&	2.04	&	52986.58	&	66	&	2511	&	76.0		&	0.9 \\
\object{HIP12961}\tablefootmark{\dagger}	&				&	M0V\tablefootmark{c}	&	9.7		&	1.64	&	52991.63	&	46	&	2226	&	46.5		&	3.0 \\
\object{HIP19394}						&				&	M3.5V\tablefootmark{d}	&	11.81	&	2.5	&	52942.80	&	35	&	2495	&	23.3		&	2.0 \\
\object{HIP38594}						&				&	M\tablefootmark{d}		&	9.96		&	1.64	&	52989.79	&	17	&	2229	&	41.1		&	2.0 \\
\object{HIP85647}\tablefootmark{\dagger}	&	GJ\,676\,A		&	M0V\tablefootmark{b}	&	9.59		&	1.85	&	53917.75	&	38	&	1520	&	36.6		&	2.4 \\
\hline
\end{tabular}
\tablefoot{
The average radial-velocity errors, $<\sigma_i(V_r)>$, are taken after removal of the planetary companion's signals.
($\dagger$) Stars with published planetary companions.
}
\tablebib{
(a)~\citet{bonfils2011} unless individually specified; 
(b)~\citet{hawley1996}; 
(c)~\citet{koen2010};
(d)~Simbad (http://simbad.u-strasbg.fr/simbad/);
(e)~\citet{esa1997}
}
\end{table*}

The sample we are using comes from the HARPS M-dwarf planet search program that started in 2003 and ended in 2009 \citep[see][]{bonfils2011}.
We used this sample in Paper I to study four known chromospheric activity indices and to select stars with long-term activity variability.
However, we now use more data that were taken in 2010 as part of the \citet{bonfils2011} program extension.
Therefore, We decided to redo the same analysis as in Paper I using the new data to detect any new cases of activity variability that could arise from more data points.

We obtained simultaneous radial-velocity, BIS, FWHM, contrast, and the \ion{Na}{i} activity indicator.
The median radial-velocity error of the nightly averaged data was 1.2 m\,s$^{-1}$.
HARPS is capable of more precise measurements \citep[e.g.][]{mayor2011} but our sample includes dim stars which make it more difficult to get high signal-to-noise ratios (SNR) and therefore higher-precision RVs.
To measure activity\footnote{From now on we will use the terms activity, \ion{Na}{i}, and sodium lines interchangeably.} we used the index based on the \ion{Na}{i} D1 and D2 lines in a same way as in Paper I \citep[see also][]{diaz2007a}.
The CCF parameters were used because of their potential as complementary long-term activity proxies \citep[e.g.][]{santos2010}.

Because of an instrumental drift detected on HARPS that affected the FWHM and contrast of the CCF, these parameters were corrected using the following expressions:
\begin{eqnarray}
FWHM_{\mathrm{corr}}= FWHM + 5.66\cdot 10^{-6} D - 1.77\cdot10^{-9} D^2 \\
contrast_{\mathrm{corr}} = contrast + 4.59\cdot10^{-5} D - 2.75\cdot10^{-8} D^2
\end{eqnarray}
where $D$ is $(BJD - 2454500)$ days. However, this correction is very small and is a long-term drift. For example the FWHM drift is only 0.1\% in five years.

Since this sample includes some stars very close to the Sun, these stars may have significant secular acceleration that could produce a trend on the observed radial-velocity.
We corrected all stars for this effect using the proper motions and parallaxes that were retrieved from the Hipparcos catalog \citep{esa1997}.
The secular accelerations were calculated following the description in \citet{zechmeister2009} and the radial-velocities subsequently corrected \citep[see also][]{bonfils2011}.

Some of the stars in this sample are planet hosts, namely Gl\,176, Gl\,433, Gl\,436, Gl\,581, Gl\,667\,C, Gl\,674, Gl\,832, Gl\,849, Gl\,876, HIP12961, and HIP85647 \citep{bonfils2011}.
Because we are only interested in variations due to activity, these planetary signals were subtracted from the observed radial-velocities by fitting keplerian functions based on the published orbital parameters.

The selection criteria was the same as in our previous paper.
We nightly averaged and binned the data into 150-day bins to average out short time-scale variability, since we are only concerned with long-term variations.
Only bins with more than three nights and stars with at least four bins were selected.
The errors used for each bin are the statistical errors on the average, $\sigma_i = \sigma/\sqrt{N}$, where $\sigma$ is the root-mean-square (rms) of the nightly averaged data in each bin, and $N$ is the number of nights included in the bin.

We also decided to select only stars with at least three years of observations, which resulted in GJ\,361, GJ\,2049, and GJ\,3218 being discarded from the sample.
This resulted in a sample of 27 stars which passed the selection criteria, whose basic parameters are presented in Table \ref{sample}. These stars have spectral types in the range M0--M5.5\,V and have $V$ magnitudes between 7.34 and 11.81. In the same table we also present the observation log, with information about the number of nights of observation, the timespan in days and years, the average signal-to-noise ratio at spectral order 56 (the order of the \ion{Na}{i} doublet, $\sim$5893\AA) and the average radial-velocity error.
The timespan of the observations range from 3.3 to 7.2 years (with a median value of 5.9 years).
The average errors on radial velocity per star for the nightly averaged data vary between 0.8 and 3.0 m\,s$^{-1}$.

\section{\ion{Na}{i} index variability} \label{vartests}
In Paper I we compared four activity indices, namely $S_{\mathrm{Ca\,II}}$, H$\alpha$, $\ion{Na}{i}$, and $\ion{He}{i}$, and arrived to the conclusion that the $\ion{Na}{i}$ index is the best to follow the long-term activity of M-dwarf stars.
We also tested our sample for variability and found that Gl\,1, Gl\,273, Gl\,433, Gl\,436, Gl\,581, Gl\,588, Gl\,667\,C, Gl\,832, Gl\,849, Gl\,877, Gl\,908, and HIP85647 showed significant ($P(F) \leq 0.05$) long-term activity variability on at least two indices.
However, we are now using the latest data with new measurements from 2010 (previously we used data covering the years 2003--2009) and consequently other stars might now show statistical variability on long time-scale activity that was not detected before.
We therefore decided to repeat the variability F-tests, this time using only the $\ion{Na}{i}$ index.

The \ion{Na}{i} activity proxy was determined as explained in Paper I, by measuring the flux in the centre of the sodium D1 and D2 lines in comparison to the flux on two reference bands.
This index is not calibrated to the bolometric flux of the stars and therefore it will depend on the effective temperature of each star (see discussion on Sect. 6.6 of Paper I).
As a consequence, this index cannot be used to compare the activity of stars with different effective temperatures.
Nevertheless, it can be used to detect activity variability with time for a given star.

In a similar fashion to what was done in Paper I, we asked whether the long time-scale variations observed in our data were of statistical significance.
As previously done in Paper I, we used the F-test with an F-value as $F = \sigma_e^2 / <\sigma_i>^2$ where $\sigma_e$ is the standard deviation and $<\sigma_i>$ the average of the error on the mean of the binned parameter for each star \citep[see][]{endl2002,zechmeister2009,bonfils2011}.
This F-test will give the probability that the observed standard deviation can be explained by the random scatter due to the internal errors.
Therefore, a low value of $P(F)$ will discriminate the stars which have significant variability not justified by the internal errors.
The results of these tests are shown in Table \ref{table:var}.
Bold values indicate probabilities lower than 0.05 (95\% significance level).
The F-value used to calculate the probabilities was $F = \sigma_e(\ion{Na}{i})^2 / <\sigma_i(\ion{Na}{i})>^2$ where $\sigma_e$ stands for the rms of the (binned) data and $<\sigma_i>$ is the average of the error on the mean.
Also shown is the number of bins $N_{bins}$, indication if the star passed the variability tests in Paper I, $Sel.$, the radial-velocity's rms $<\sigma_e(V_r)>$ and average error on the mean $<\sigma_i(V_r)>$, and the average value of the \ion{Na}{i} index, $<\ion{Na}{i}>$.

Fourteen out of the 27 stars show long-term activity variability with $P(F) \leq 0.05$, representing 52\% of the sample.
These stars are Gl\,273, Gl\,433, Gl\,436, Gl\,526, Gl\,581, Gl\,588, Gl\,667\,C, Gl\,699, Gl\,832, Gl\,849, Gl\,876, Gl\,877, Gl\,908, and HIP85647.
These are the same stars that passed the tests in Paper I plus Gl\,526, Gl\,699, and Gl\,876, and less Gl\,1.
Four more stars, Gl\,680, Gl\,887, HIP19394, and HIP38594, have $0.05 < P(F) \leq 0.1$, which if they are considered increases the percentage of stars with variability to 67\%.
In Paper I we found 17 stars out of 30 showing variability in the \ion{Na}{i} index with $P(F) \leq 0.1$, representing 57 \% of the sample.
This represents an increase of variable stars by 10\% from our previous work just by adding one more year of observations, and thus, more of these stars could show long-term variability if a longer time-span is considered.
We should note that Gl\,1 passed the variability test on Paper I but not here.
This is because on Paper I this star passed the test due to its variability in the $S_{\ion{Ca}{ii}}$ and H$\alpha$ indices, while not showing significant variability in \ion{Na}{i}.
Since we used only this index now and Gl\,1 have not had more data points in 2010, this star now failed to pass the F-test based just on \ion{Na}{i}.

By using a sample of old FGK stars in the solar neighbourhood studied for up to 7 years, \citet{lovis2011} found that 61\% have a magnetic activity cycle.
Although we cannot confirm the cyclic nature of our detected long-term activity variations, our results are compatible with those of \citet{lovis2011}.

From now on we will consider only these 14 stars for the rest of the study, unless specified.

\begin{table*}[tbp] 
\caption{Statistics for RV and \ion{Na}{i}, and probabilities $P(F)$ of the variability F-tests for activity. All parameters were calculated for the data with 150 day bins.}
\label{table:var}
\centering
\begin{tabular}{l c c c c | c c c c} \\
\hline
\hline
\multicolumn{1}{l}{Star} &
\multicolumn{1}{c}{$N_{bins}$} &
\multicolumn{1}{c}{Sel.} &
\multicolumn{1}{c}{$\sigma_e(V_r)$} &
\multicolumn{1}{c}{$<\sigma_i(V_r)>$} &
\multicolumn{1}{c}{$<\ion{Na}{i}>$} &
\multicolumn{1}{c}{$\sigma_e(\ion{Na}{i})$} &
\multicolumn{1}{c}{$<\sigma_i(\ion{Na}{i})>$} &
\multicolumn{1}{c}{$P(F)$} \\
\multicolumn{1}{l}{} &
\multicolumn{1}{c}{} &
\multicolumn{1}{c}{} &
\multicolumn{1}{c}{[m\,s$^{-1}$]} &
\multicolumn{1}{c}{[m\,s$^{-1}$]} \\
\hline
Gl\,1		& 5 & Yes	&	0.58	&	0.71	&	0.104	&	0.0021	&	0.0013	&	0.18		 \\
Gl\,176	& 7 & No	&	2.7	&	0.94	&	0.188	&	0.0074	&	0.0044	&	0.12		 \\
Gl\,205	& 6 & No	&	2.9	&	0.82	&	0.179	&	0.0046	&	0.0029	&	0.16		 \\
Gl\,273	& 6 & Yes	&	1.5	&	0.49	&	0.075	&	0.0037	&	0.0016	&	\textbf{0.042}	 \\
Gl\,382	& 5 & No	&	2.1	&	2.2	&	0.196	&	0.0052	&	0.0027	&	0.12		 \\
Gl\,393	& 5 & No	&	1.6	&	0.71	&	0.146	&	0.0037	&	0.0023	&	0.18		 \\
Gl\,433	& 6 & Yes	&	1.2	&	0.71	&	0.141	&	0.0034	&	0.0011	&	\textbf{0.011}	 \\
Gl\,436	& 8 & Yes	&	0.43	&	0.52	&	0.100	&	0.0042	&	0.0012	&	\textbf{0.0017}	 \\
Gl\,479	& 4 & No	&	2.0	&	0.97	&	0.186	&	0.0040	&	0.0029	&	0.31		 \\
Gl\,526	& 4 & No	&	1.8	&	0.69	&	0.158	&	0.0035	&	0.0011	&	\textbf{0.042}	 \\
Gl\,551	& 6 & No	&	1.5	&	0.83	&	0.493	&	0.057	&	0.044	&	0.28		 \\
Gl\,581	& 13 & Yes &	0.96	&	0.51	&	0.071	&	0.0026	&	0.0013	&	\textbf{0.012}	 \\
Gl\,588	& 5 & Yes	&	1.3	&	0.39	&	0.131	&	0.0059	&	0.0010	&	\textbf{0.0025}	 \\
Gl\,667\,C	& 8 & Yes	&	0.73	&	0.61	&	0.096	&	0.0030	&	0.0014	&	\textbf{0.031}	 \\
Gl\,674	& 5 & No	&	2.3	&	0.73	&	0.138	&	0.0037	&	0.0029	&	0.32		 \\
Gl\,680	& 6 & No	&	5.9	&	0.93	&	0.146	&	0.0048	&	0.0022	&	0.059		 \\
Gl\,699	& 5 & No	&	1.5	&	0.61	&	0.059	&	0.0082	&	0.0027	&	\textbf{0.027}	 \\
Gl\,832	& 5 & Yes	&	0.30	&	0.66	&	0.132	&	0.0039	&	0.0015	&	\textbf{0.047}	 \\
Gl\,849	& 6 & Yes	&	0.29	&	0.77	&	0.124	&	0.0032	&	0.0013	&	\textbf{0.032}	 \\
Gl\,876	& 6 & No	&	2.3	&	1.8	&	0.081	&	0.0047	&	0.0017	&	\textbf{0.020}	 \\
Gl\,877	& 5 & Yes	&	2.7	&	1.3	&	0.106	&	0.0074	&	0.0020	&	\textbf{0.014}	 \\
Gl\,887	& 4 & No	&	1.6	&	1.1	&	0.173	&	0.0038	&	0.0014	&	0.062		 \\
Gl\,908	& 9 & Yes	&	1.2	&	0.55	&	0.131	&	0.0031	&	0.0014	&	\textbf{0.020}	 \\
HIP12961	& 6 & No &	1.3	&	0.89	&	0.172	&	0.0028	&	0.0016	&	0.12		 \\
HIP19394	& 7 & No &	5.3	&	2.0	&	0.094	&	0.0033	&	0.0019	&	0.096		 \\
HIP38594	& 4 & No &	2.6	&	1.2	&	0.169	&	0.0058	&	0.0020	&	0.055		 \\
HIP85647	& 6 & Yes &	1.5	&	1.3	&	0.145	&	0.0041	&	0.0017	&	\textbf{0.039}	 \\
\hline
\end{tabular}
\end{table*}

\section{Activity cycle fits} \label{cycles_fit}
Some stars show clear long-term variation of their activity, well visible with a general trend (increase or decrease of activity during several observational seasons).
These include stars that have not passed the variability F-tests of Sect. \ref{vartests}.
There is evidence that the periodic activity cycles of M-dwarf stars can be well fitted by a sinusoidal function \citep[see][]{cincunegui2007a,diaz2007b,buccino2011}. Given the usually reduced number of points in our study, fitting other functions (e.g. Keplerian) would not bring further constraints since it would increase the number of free fitting parameters close to the number of data points.
We therefore tried to fit a sinusoidal signal to \ion{Na}{i} data to obtain the periods of such cycles (minimum periods for the cases of data covering less than one period).
The fitted sinusoidal signal was of the form
\begin{eqnarray}
\ion{Na}{i}_{sine} = A \sin \left( \frac{2\pi}{P_{cycle}}  t + \phi \right) + \delta,
\end{eqnarray}
where $A$ is the semi-amplitude in units of \ion{Na}{i}, $P_{cycle}$ the activity cycle period in days, $t$ the observation epoch in days, $\phi$ the phase, and $\delta$ a constant offset.
The quality of the fit was obtained by a p-value, which gives the probability that the data is better fit by a straight line than by a sinusoidal signal.
This p-value was calculated via an F-value given by
\begin{eqnarray}
F(\chi^2) = (N - 4) \, \frac{\chi^2_{line} - \chi^2_{sine}}{\chi^2_{sine}},
\end{eqnarray}
where $line$ and $sine$ identify if the model used was the linear fit or sinusoidal fit, respectively, and $N$ is the number of data points.
The chi-squared function used was
\begin{eqnarray}
\chi^2 = \sum^N_1 \left[ \frac{\ion{Na}{i}_n - \ion{Na}{i}_{n,sine}}{\sigma_i(\ion{Na}{i})_n} \right]^2,~\mathrm{for}~1 \leq n \leq N,
\end{eqnarray} 
where $\ion{Na}{i}_n$ are the observed values of the activity index, $\ion{Na}{i}_{n,sine}$ are the activity values obtained from the fitted $sine$ model, and $\sigma_i(\ion{Na}{i})_n$ the internal errors of each individual $\ion{Na}{i}_n$.
Smaller p-values will indicate that the data was better fitted by a sinusoid.
Note that these fits were carried out without taking in consideration whether the star passed the variability F-tests in Sect. \ref{vartests} or not.
Also, it is only possible to calculate the probabilities for the stars which have five or more data points, otherwise $(N - 4) \leq 0$.
Eight stars had their long-term activity successfully fitted by a sinusoidal signal, with $P(\chi^2) \leq 0.1$. These stars are Gl\,1, Gl\,382, Gl\,433, Gl\,581, Gl\,667\,C, Gl\,680, Gl\,832, and HIP85647.
However, Gl\,1, Gl\,382, and Gl\,680 have not passed the long-term variability F-tests and therefore they will be discarded from discussion.
The others will be discussed individually in Sect. \ref{indstars}.

\section{Radial-velocity, activity and CCF parameters}\label{rv_s_ccf}

\onlfig{1}{
\begin{figure*}[tbp]
\begin{center}
\resizebox{\hsize}{!}{\includegraphics{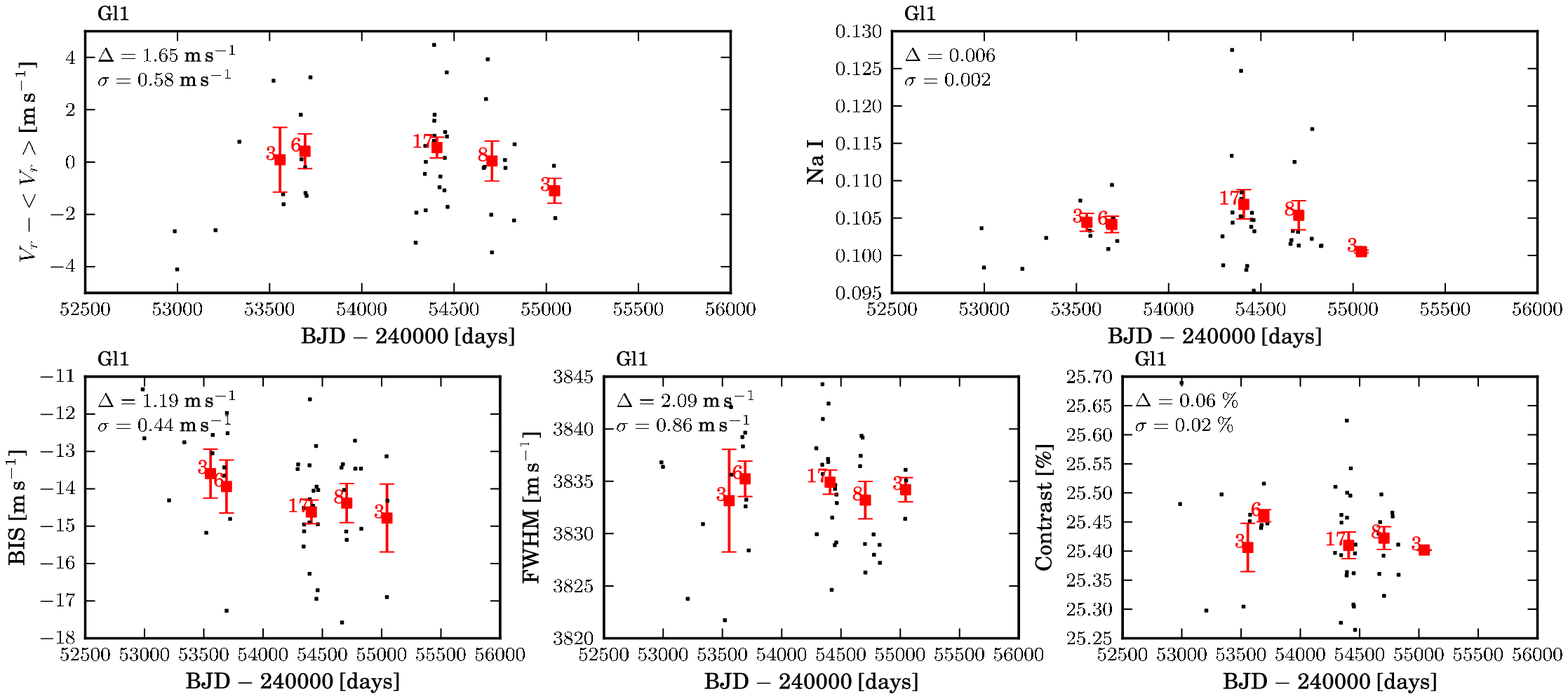}}
\resizebox{\hsize}{!}{\includegraphics{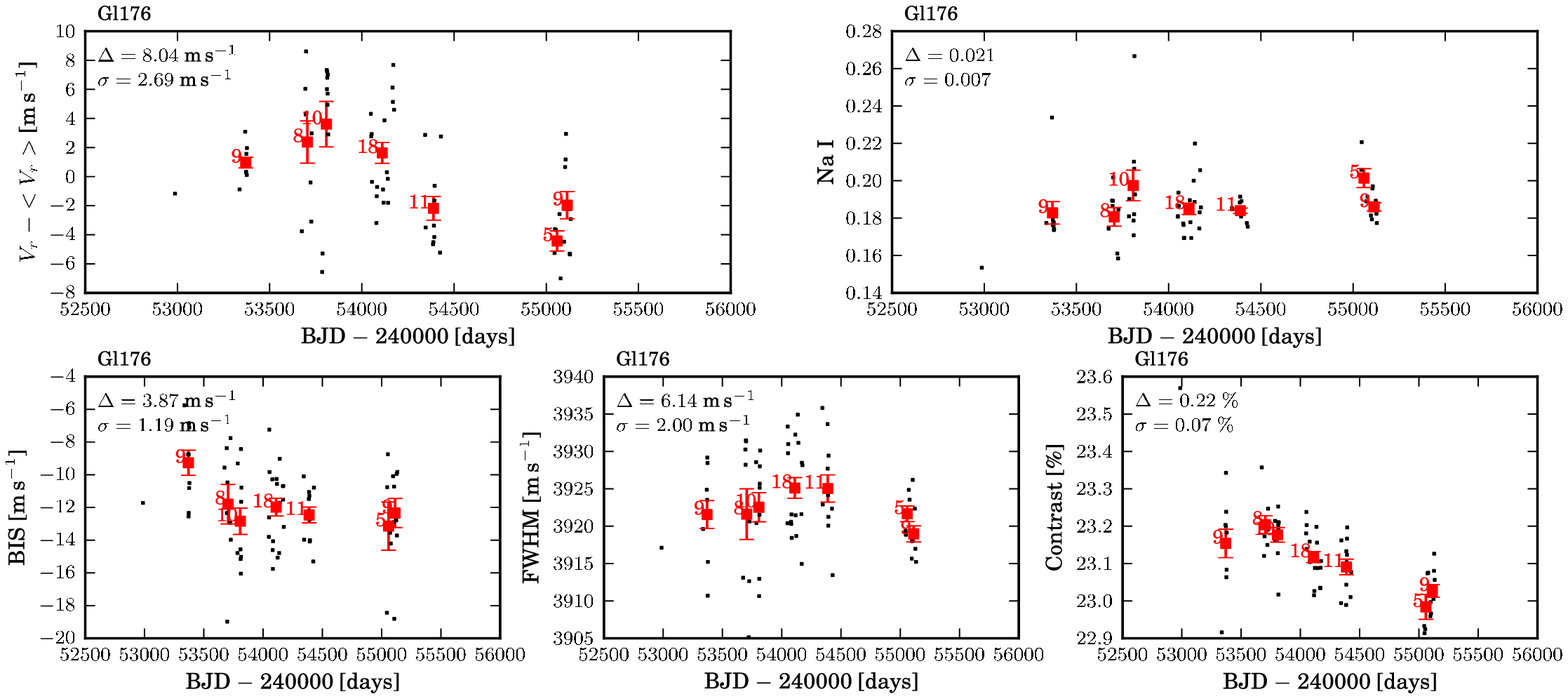}}
\resizebox{\hsize}{!}{\includegraphics{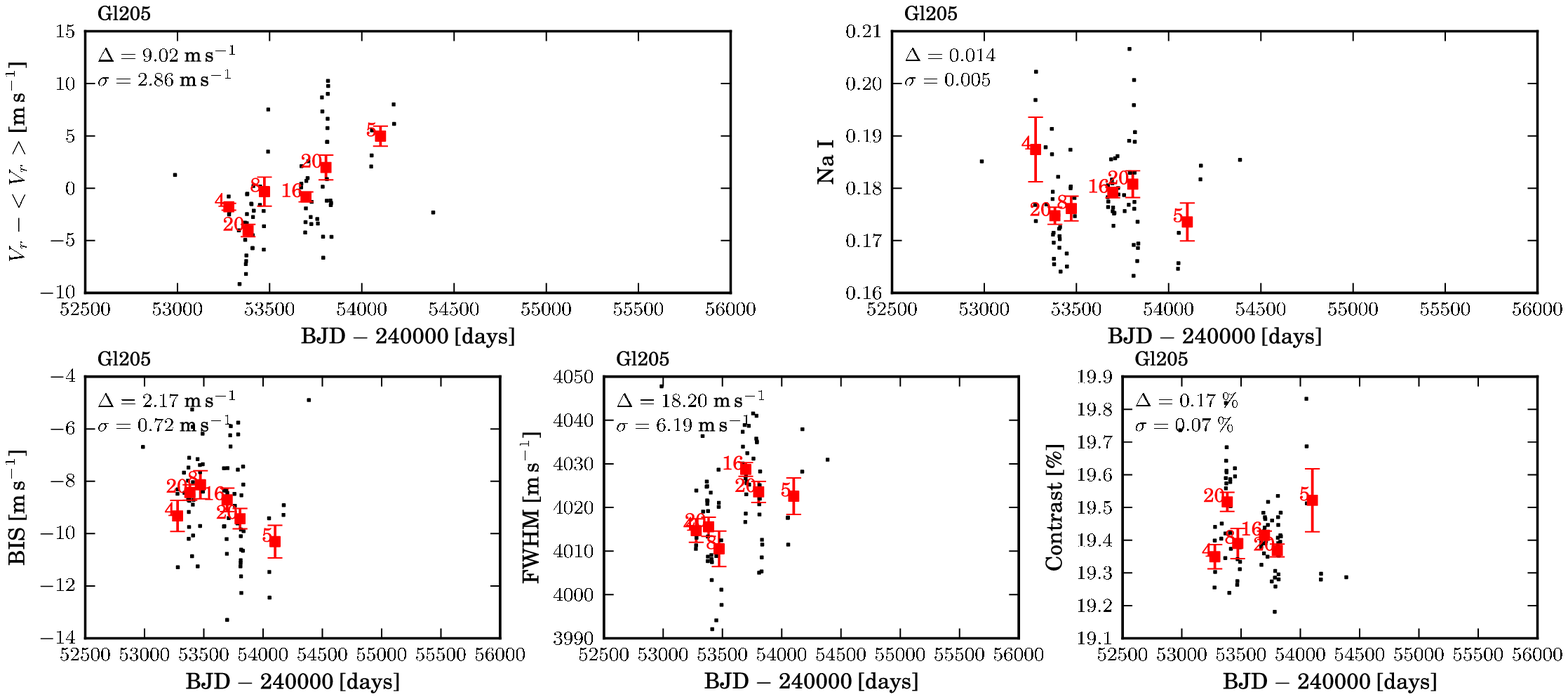}}
\caption{Time-series of radial-velocity, \ion{Na}{i} index, and BIS, FWHM, and contrast of the CCF line profile.}
\label{all_plots}
\end{center}
\end{figure*}

\begin{figure*}[tbp]
\ContinuedFloat
\begin{center}
\resizebox{\hsize}{!}{\includegraphics{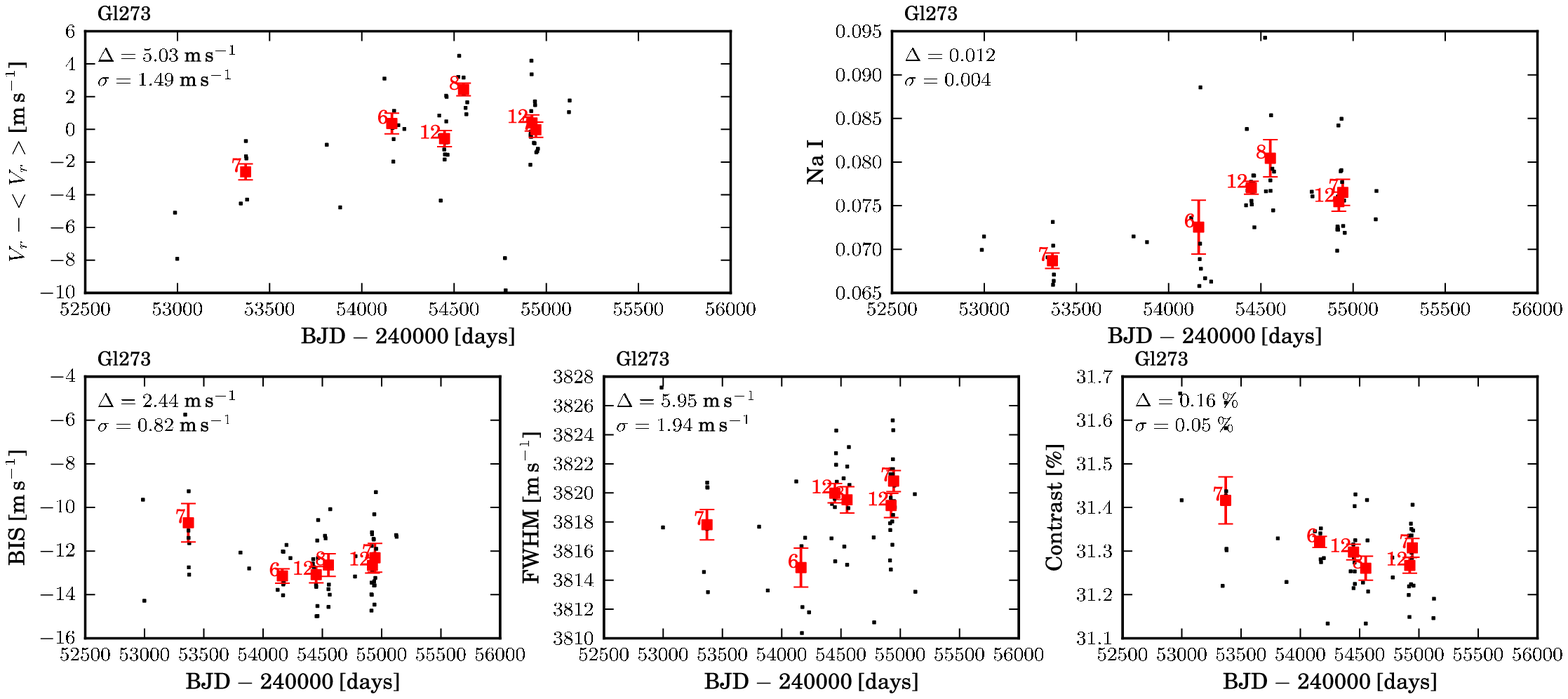}}
\resizebox{\hsize}{!}{\includegraphics{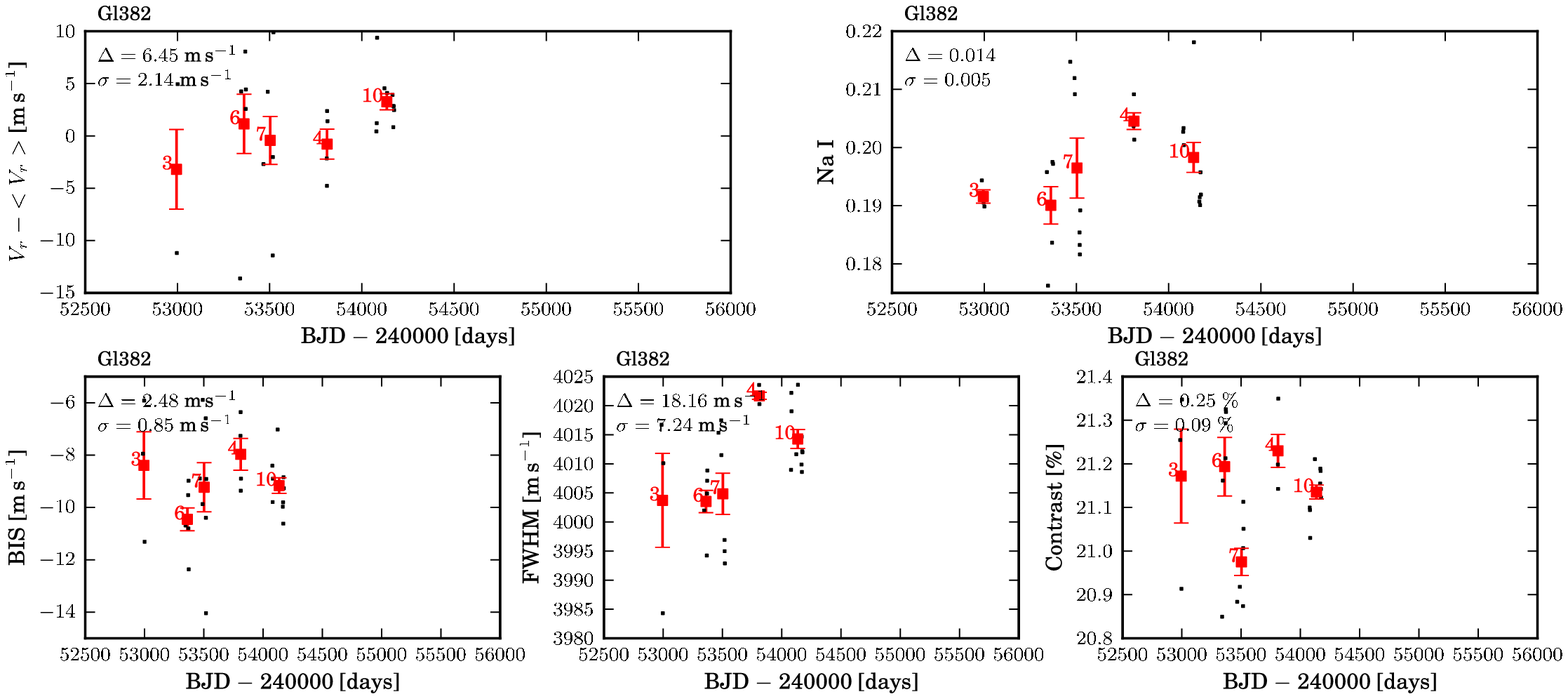}}
\resizebox{\hsize}{!}{\includegraphics{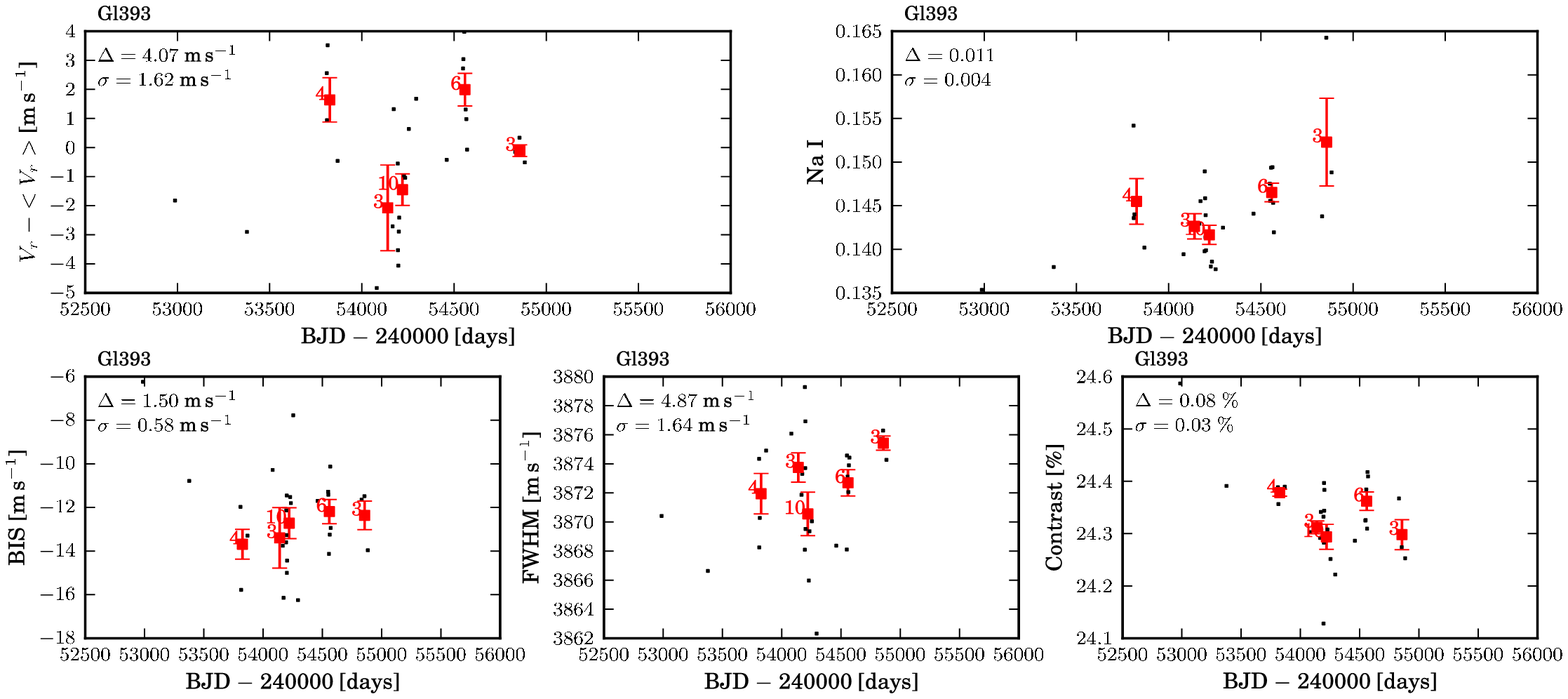}}
\caption{Continued.}
\end{center}
\end{figure*}

\begin{figure*}[tbp]
\ContinuedFloat
\begin{center}
\resizebox{\hsize}{!}{\includegraphics{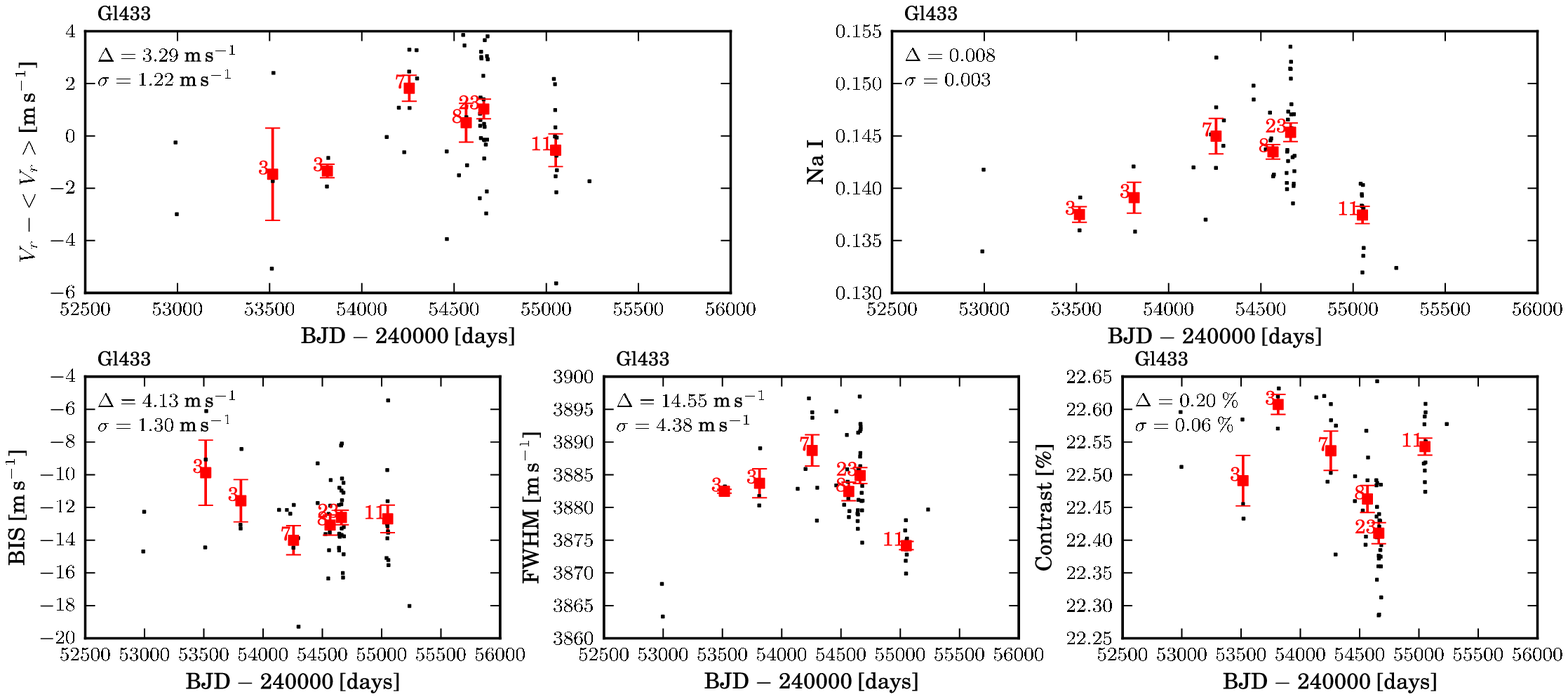}}
\resizebox{\hsize}{!}{\includegraphics{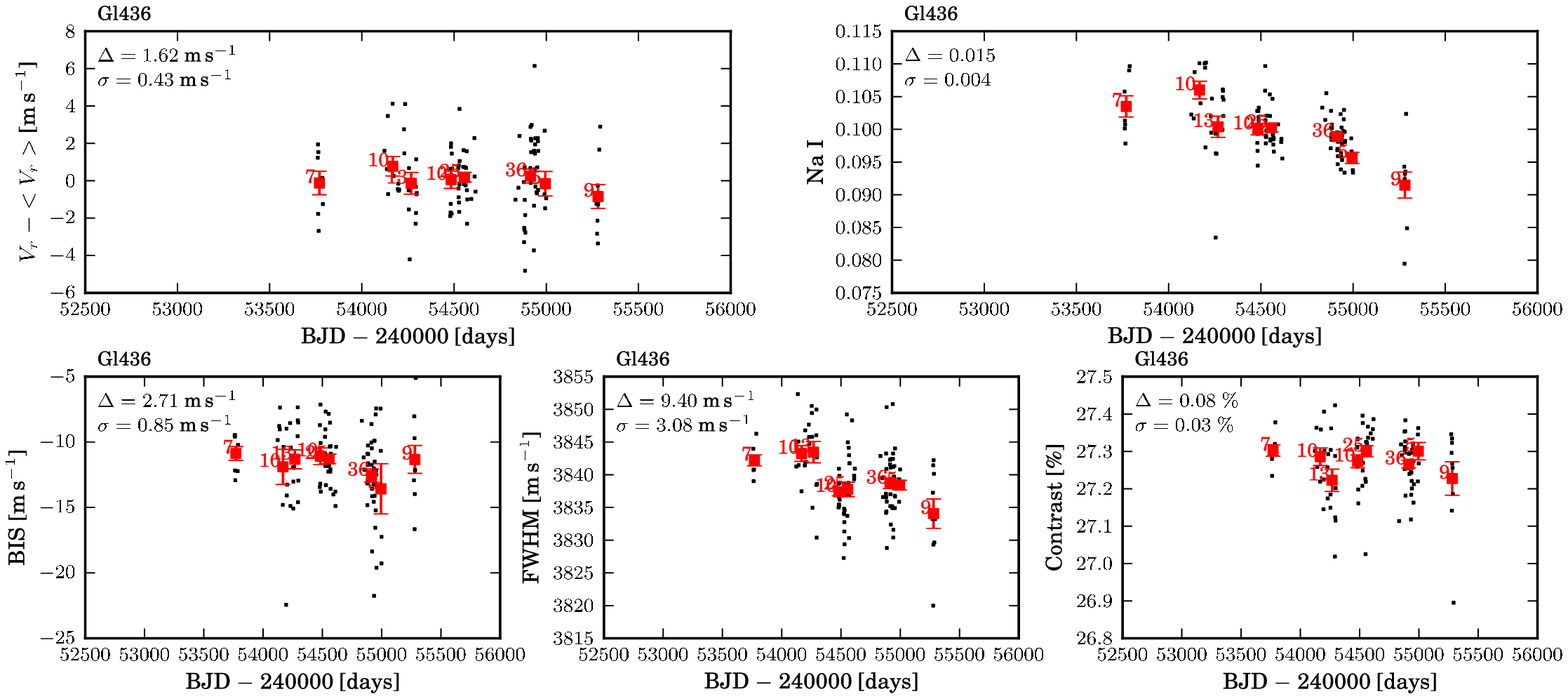}}
\resizebox{\hsize}{!}{\includegraphics{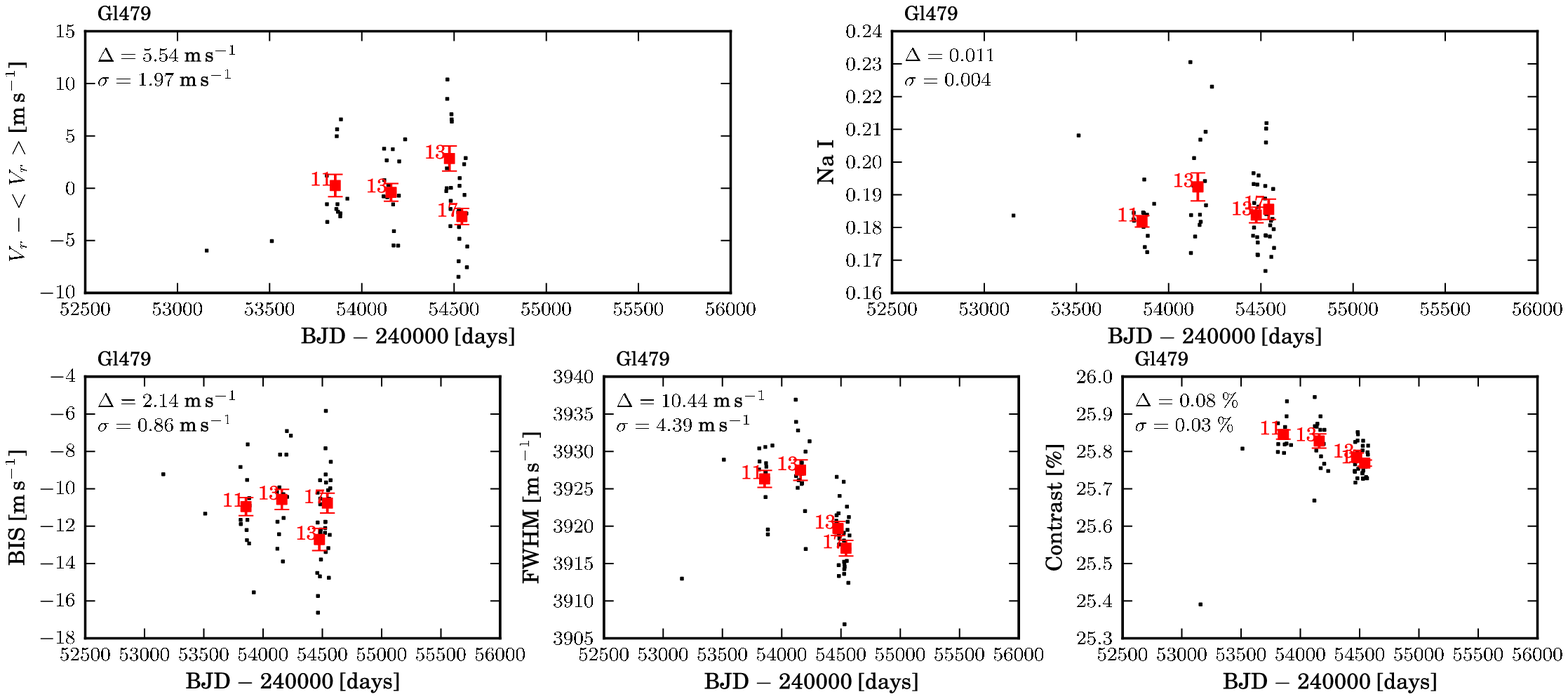}}
\caption{Continued.}
\end{center}
\end{figure*}

\begin{figure*}[tbp]
\ContinuedFloat
\begin{center}
\resizebox{\hsize}{!}{\includegraphics{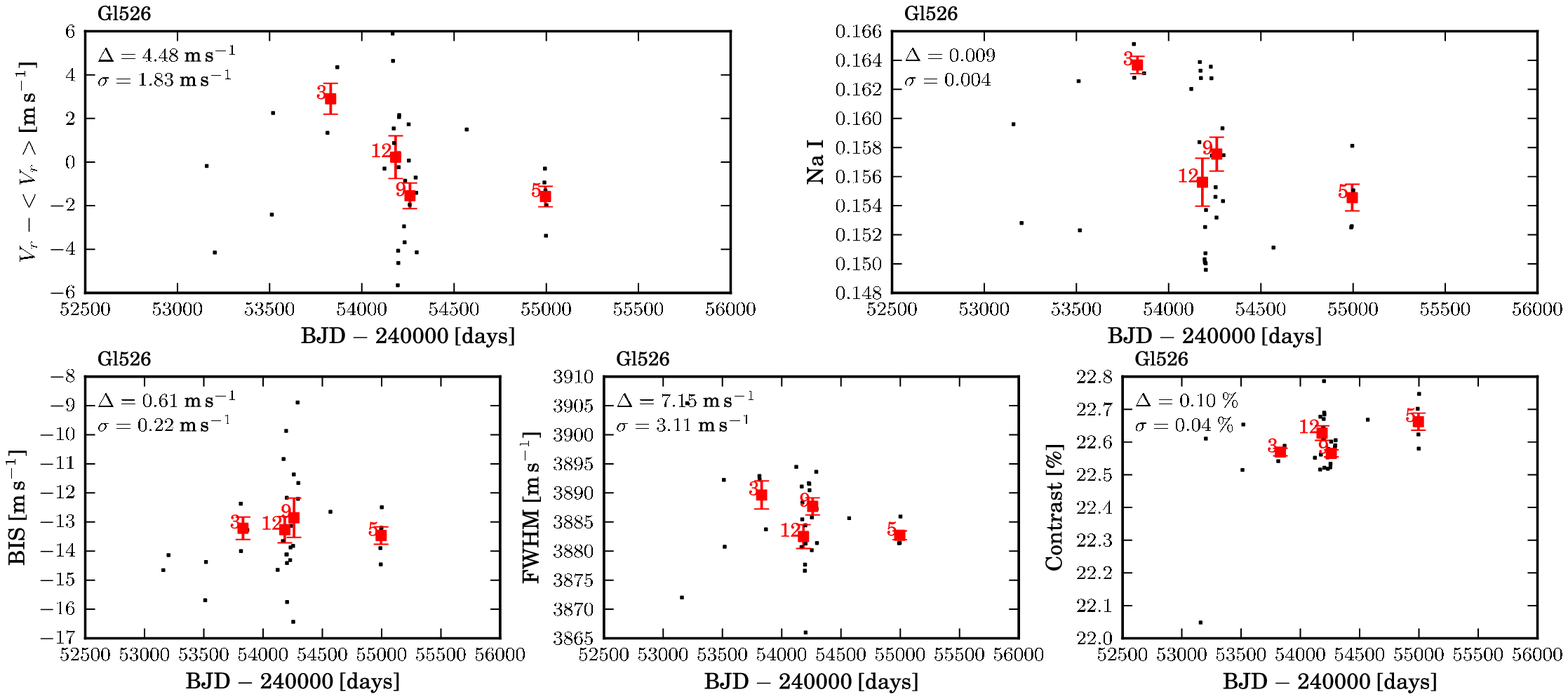}}
\resizebox{\hsize}{!}{\includegraphics{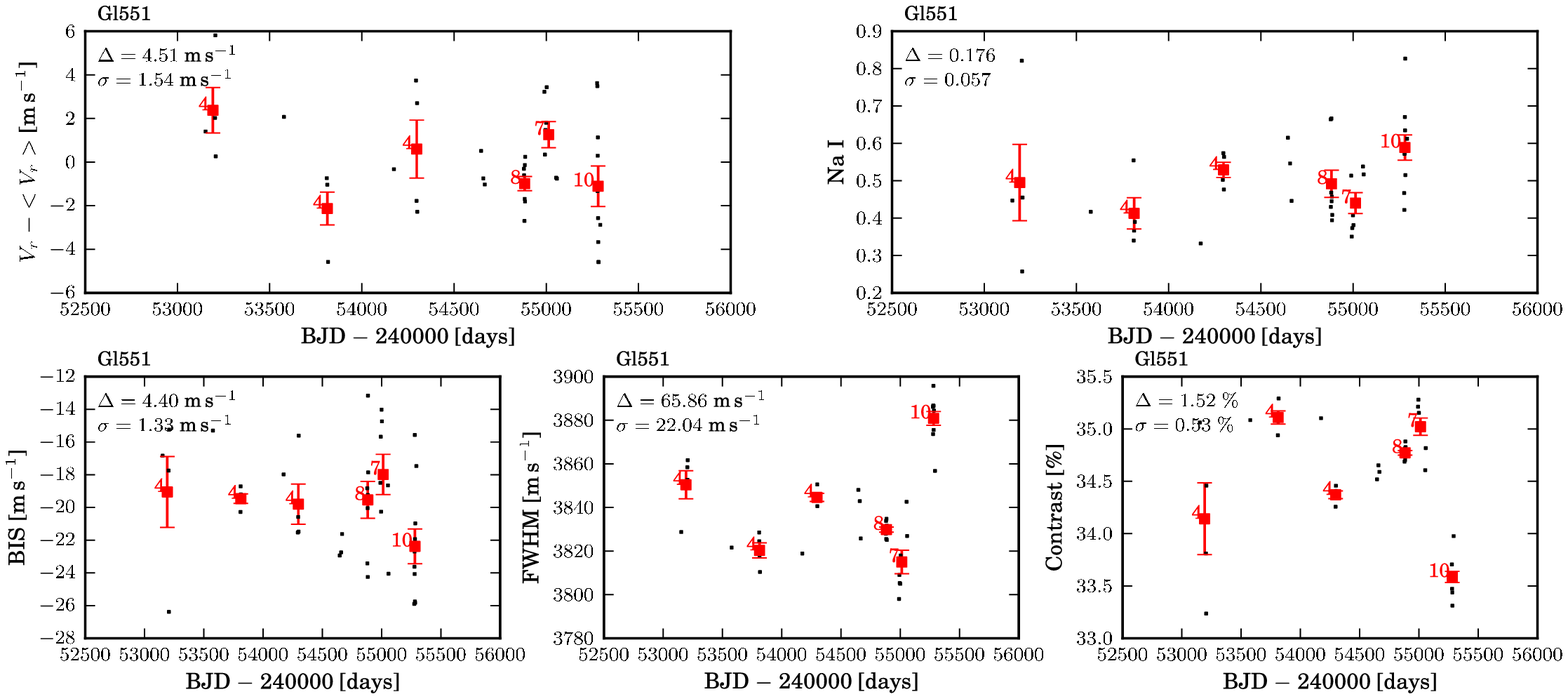}}
\resizebox{\hsize}{!}{\includegraphics{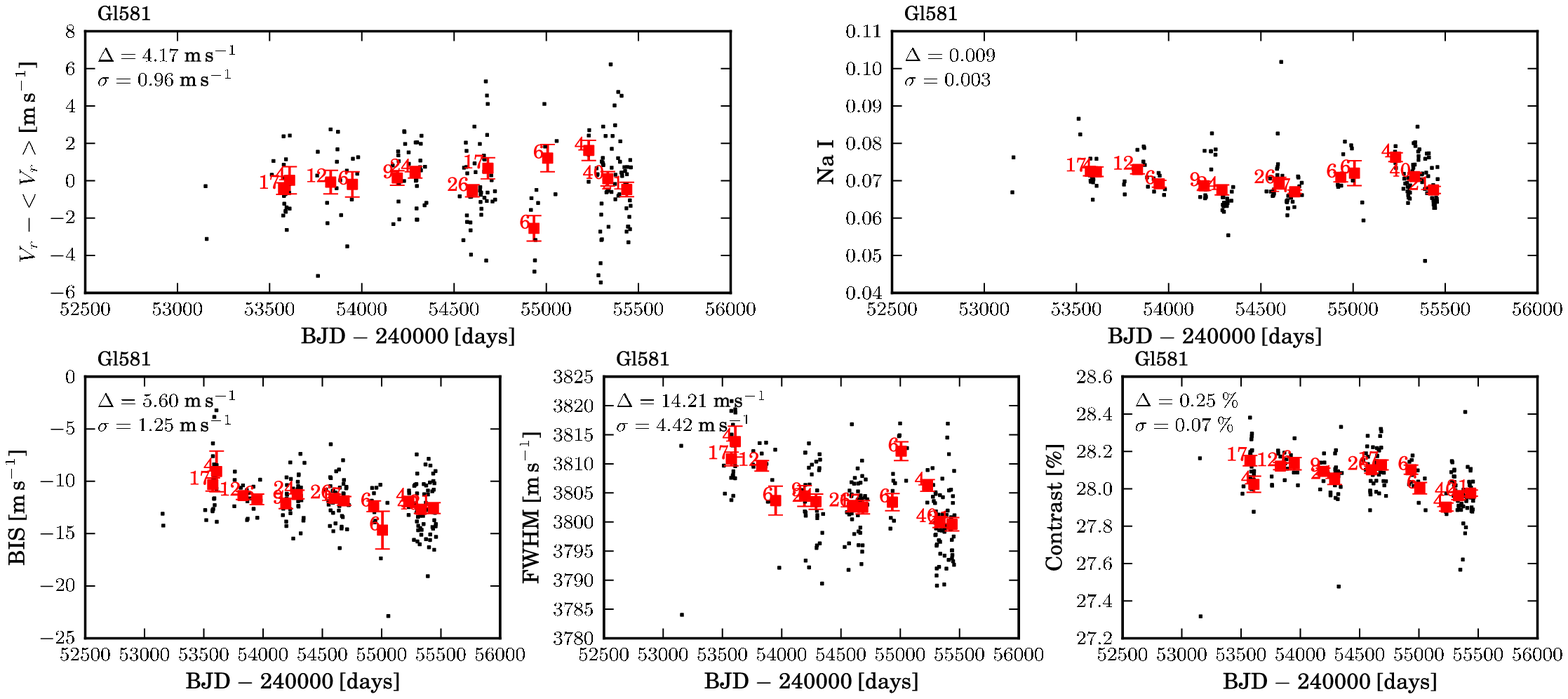}}
\caption{Continued.}
\end{center}
\end{figure*}

\begin{figure*}[tbp]
\ContinuedFloat
\begin{center}
\resizebox{\hsize}{!}{\includegraphics{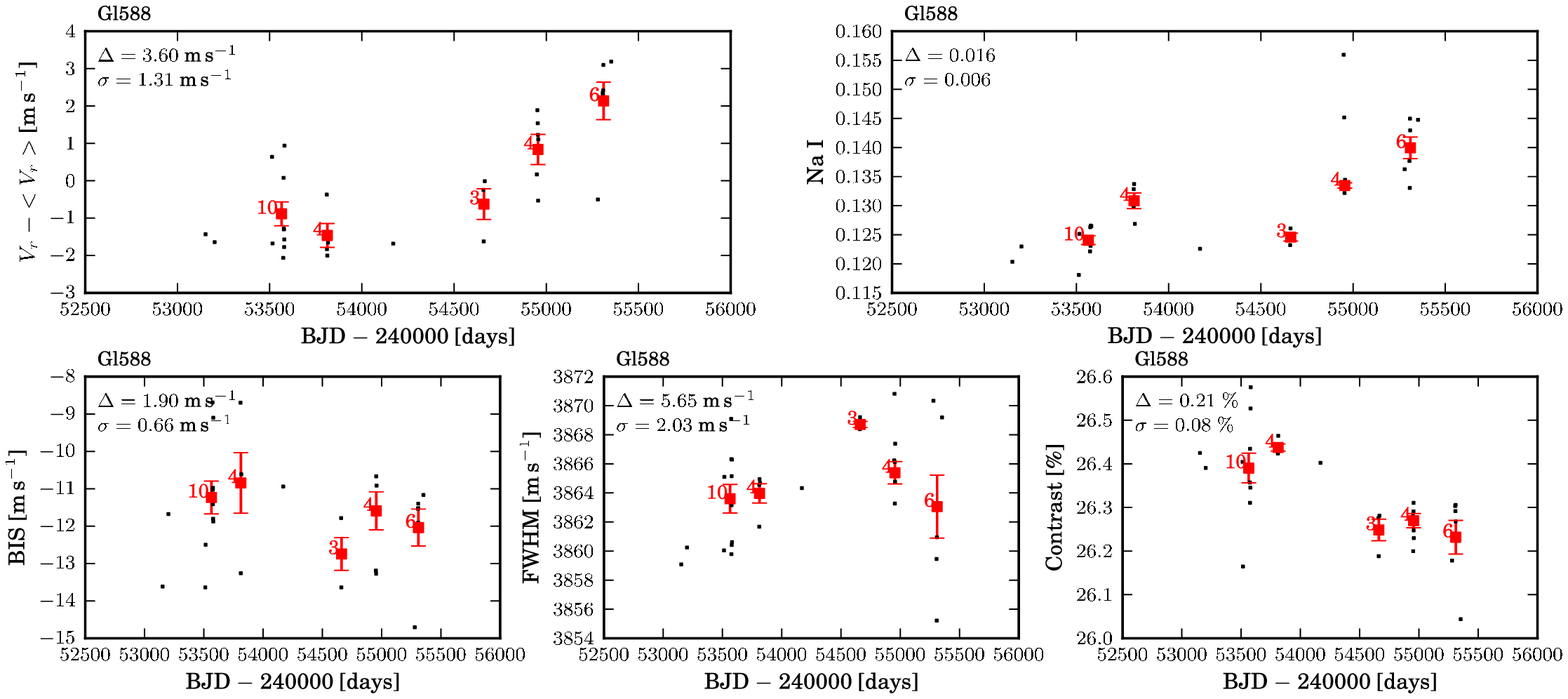}}
\resizebox{\hsize}{!}{\includegraphics{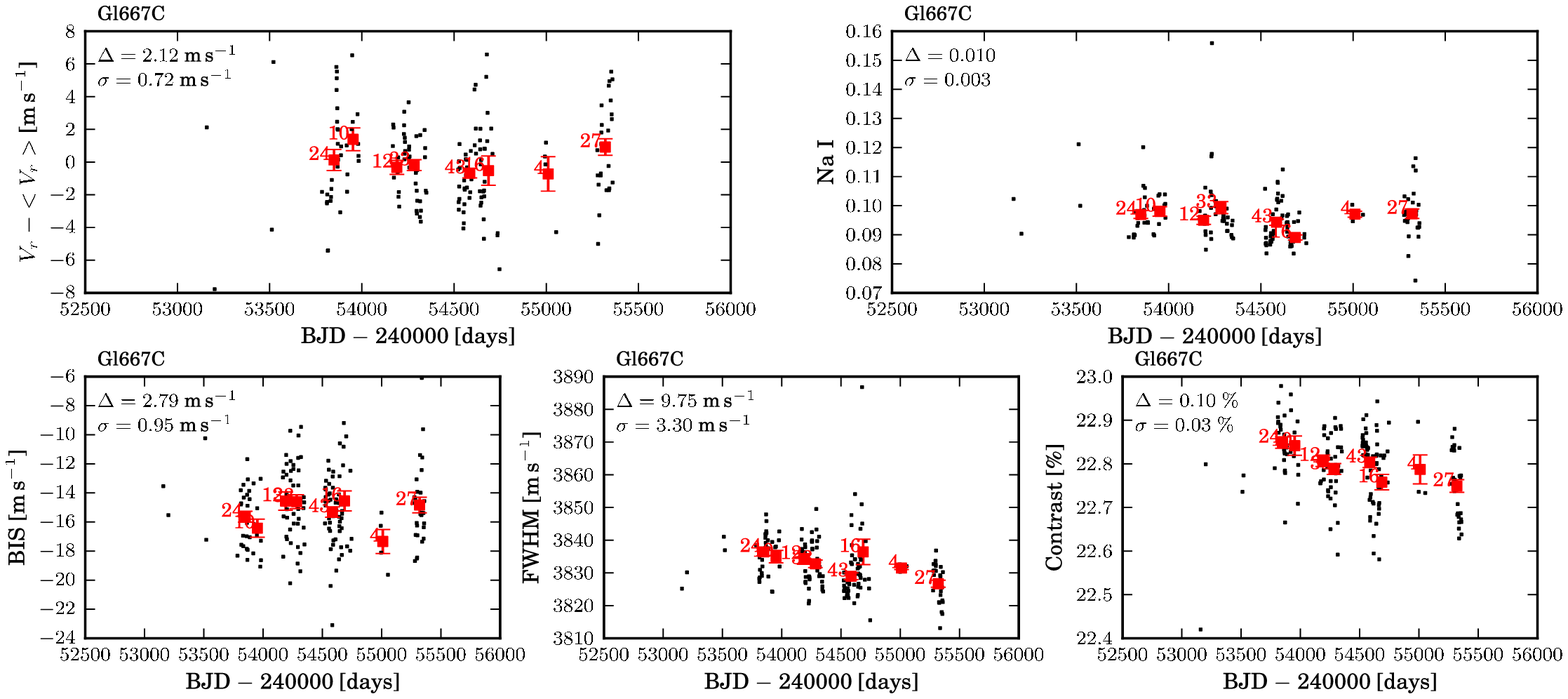}}
\resizebox{\hsize}{!}{\includegraphics{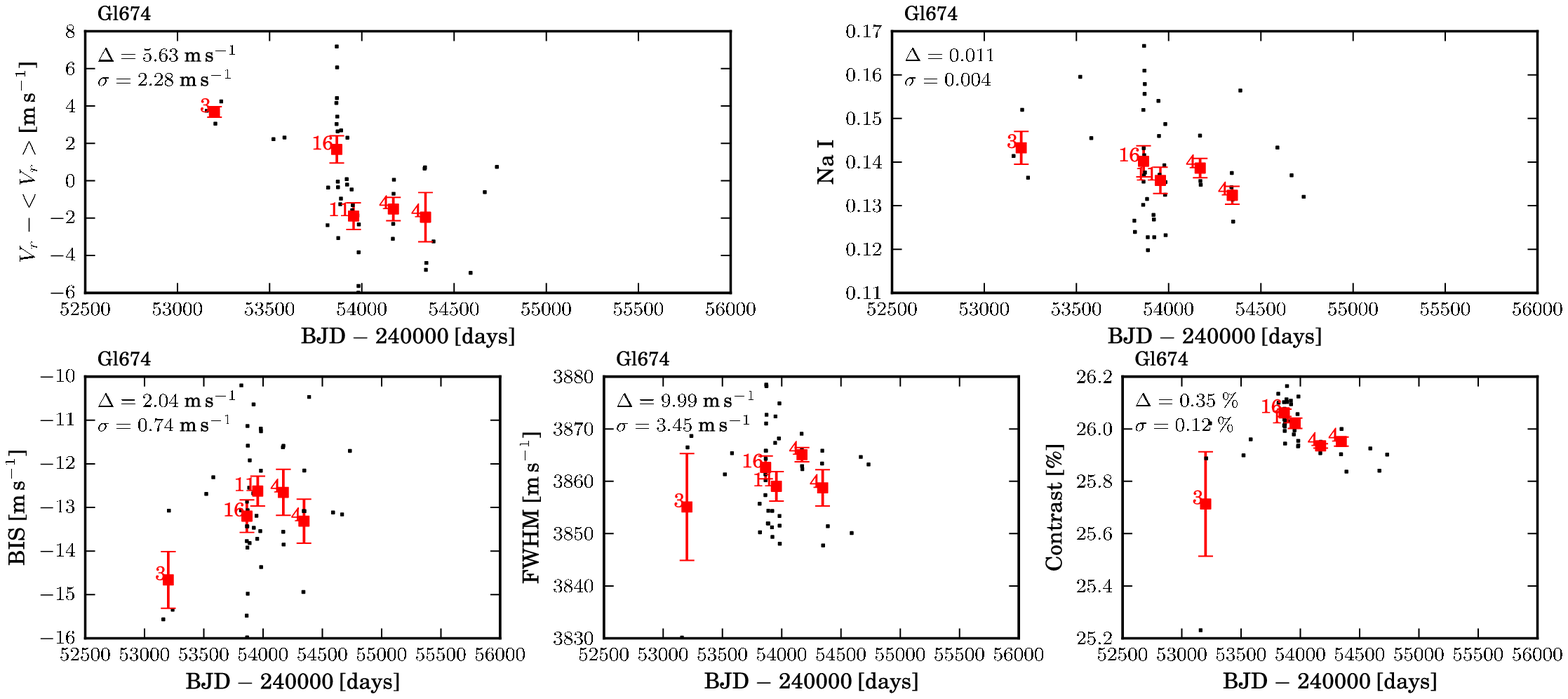}}
\caption{Continued.}
\end{center}
\end{figure*}

\begin{figure*}[tbp]
\ContinuedFloat
\begin{center}
\resizebox{\hsize}{!}{\includegraphics{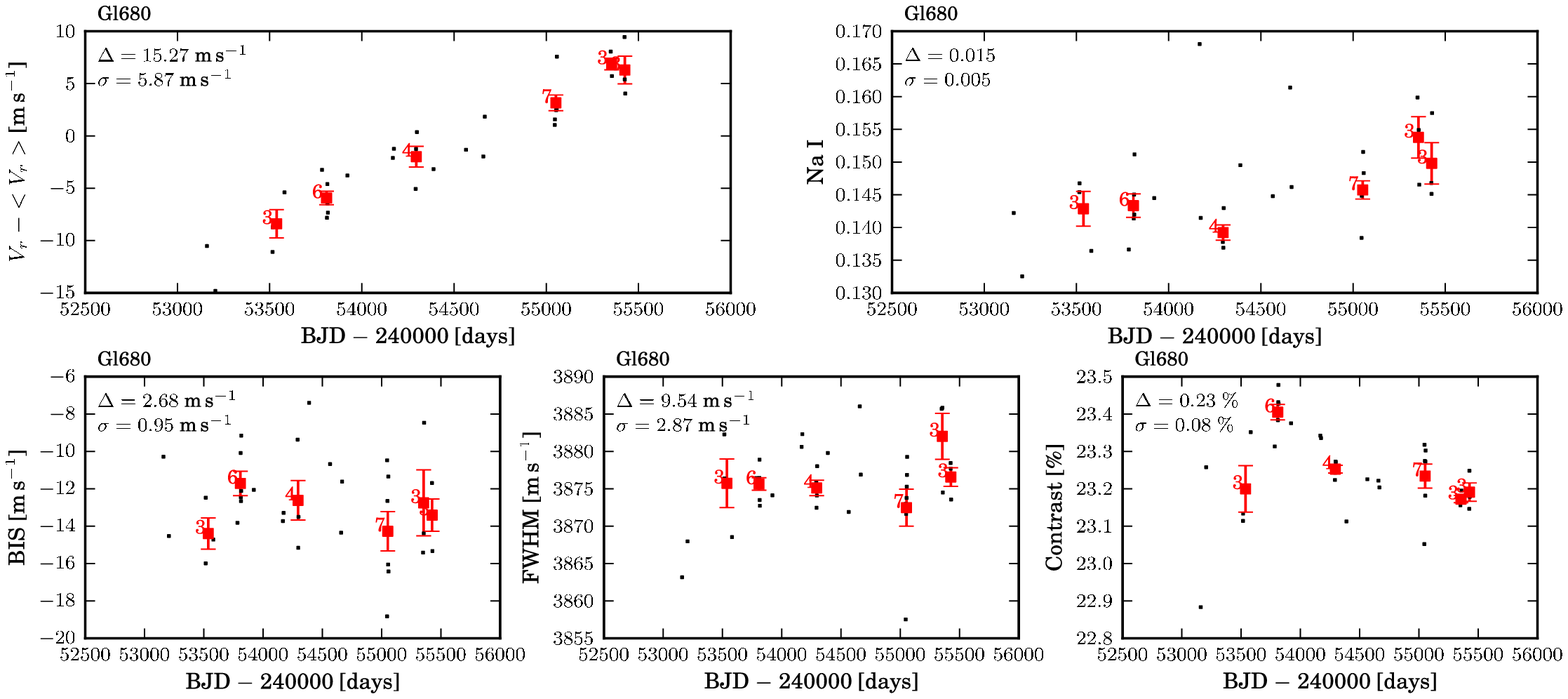}}
\resizebox{\hsize}{!}{\includegraphics{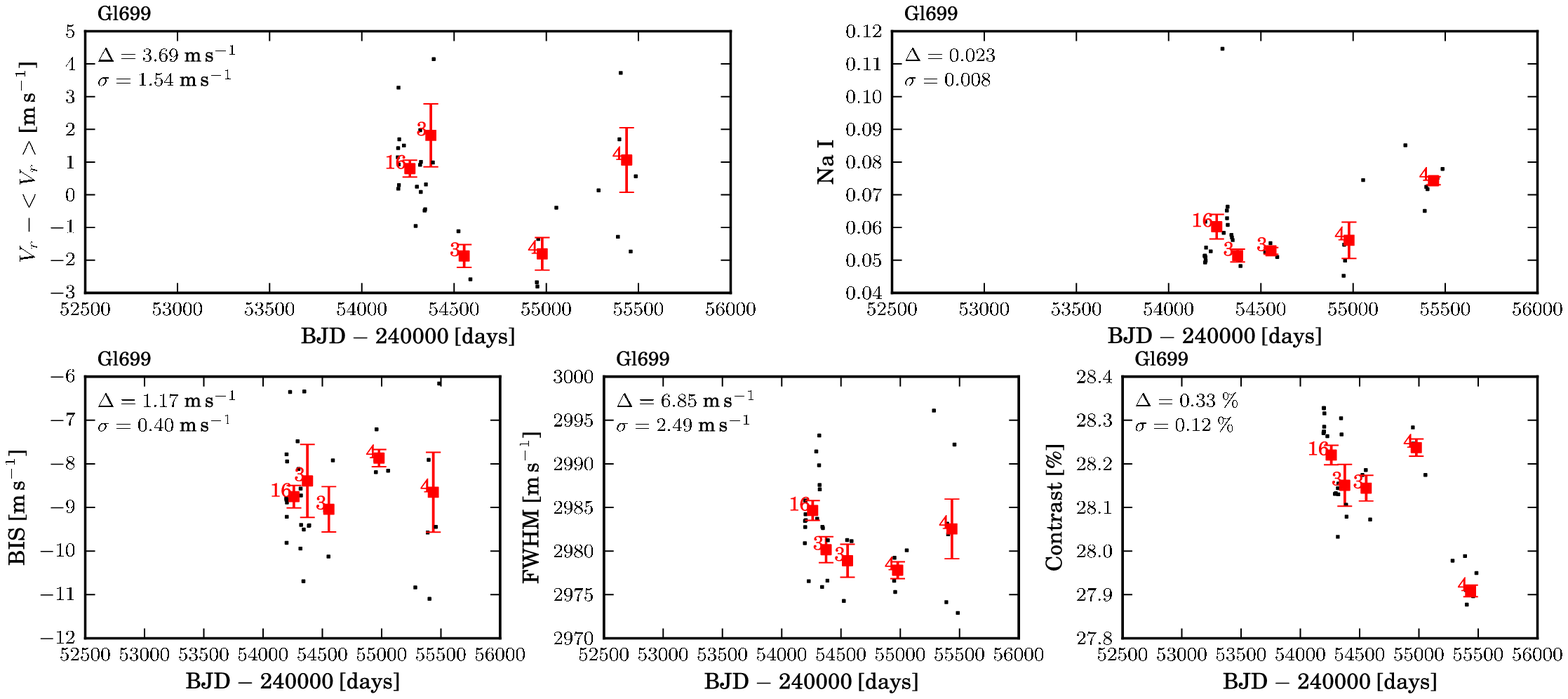}}
\resizebox{\hsize}{!}{\includegraphics{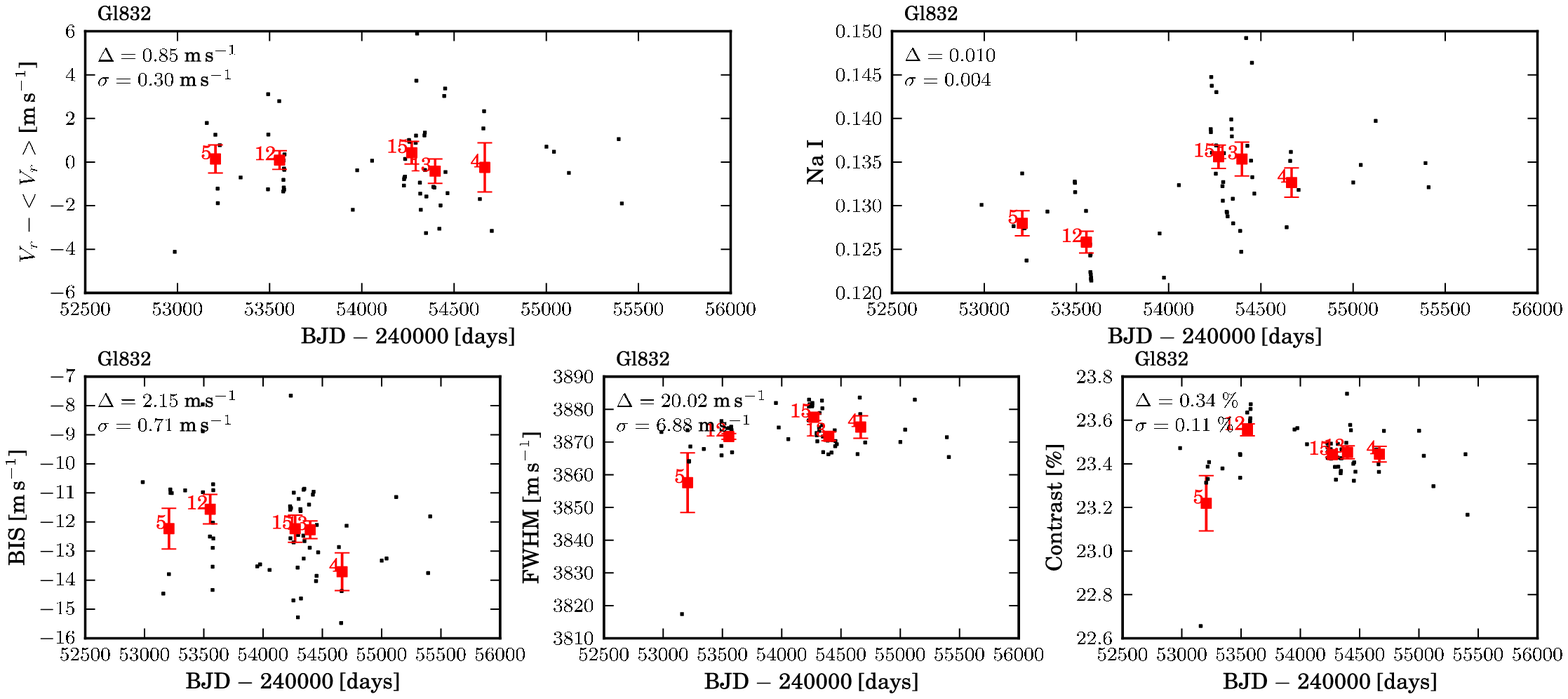}}
\caption{Continued.}
\end{center}
\end{figure*}

\begin{figure*}[tbp]
\ContinuedFloat
\begin{center}
\resizebox{\hsize}{!}{\includegraphics{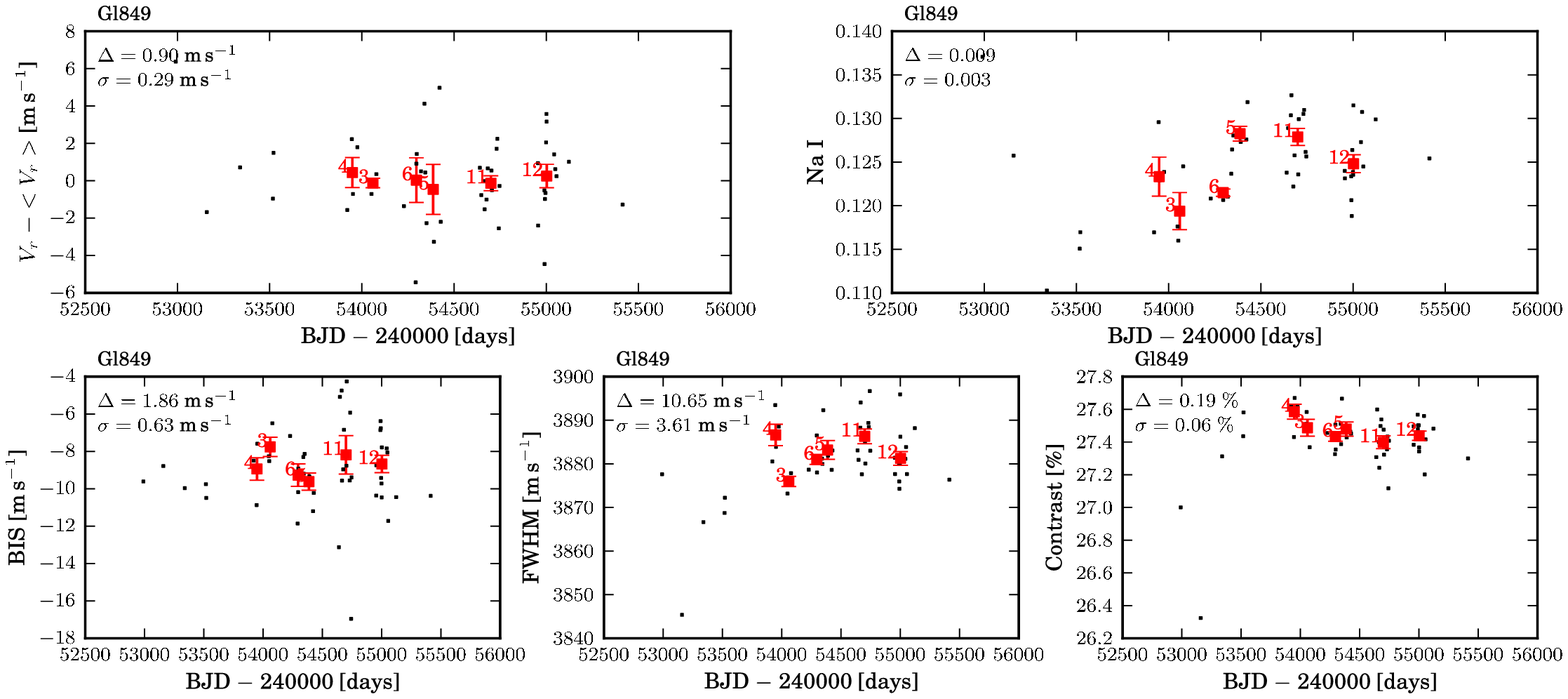}}
\resizebox{\hsize}{!}{\includegraphics{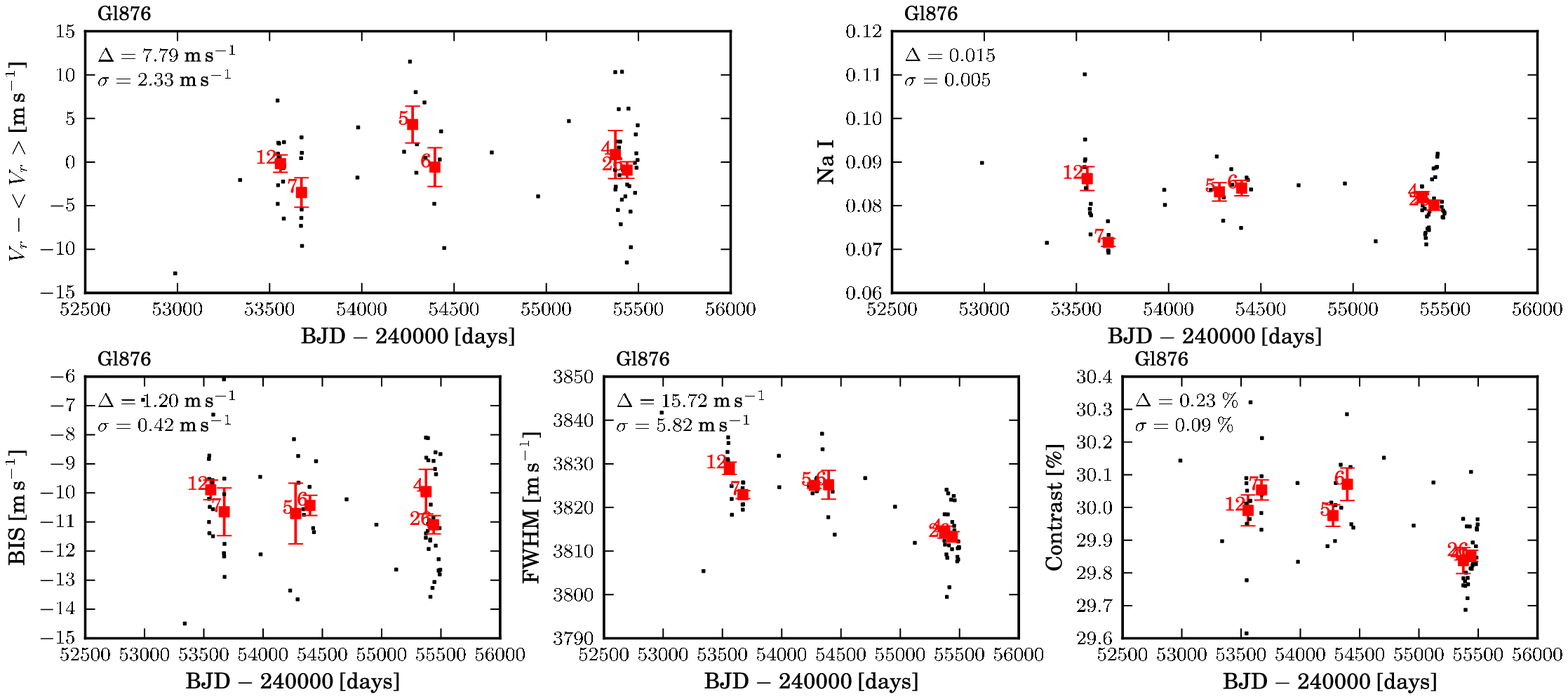}}
\resizebox{\hsize}{!}{\includegraphics{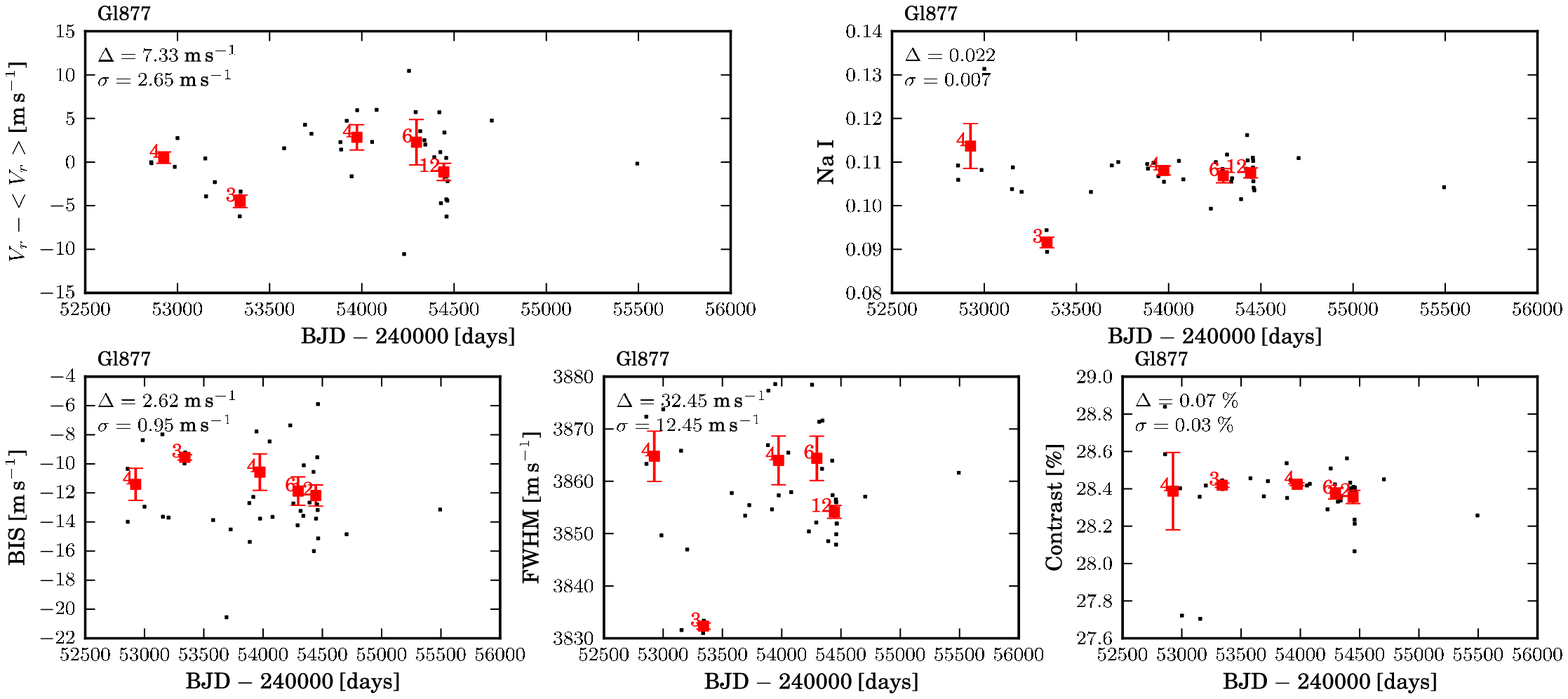}}
\caption{Continued.}
\end{center}
\end{figure*}

\begin{figure*}[tbp]
\ContinuedFloat
\begin{center}
\resizebox{\hsize}{!}{\includegraphics{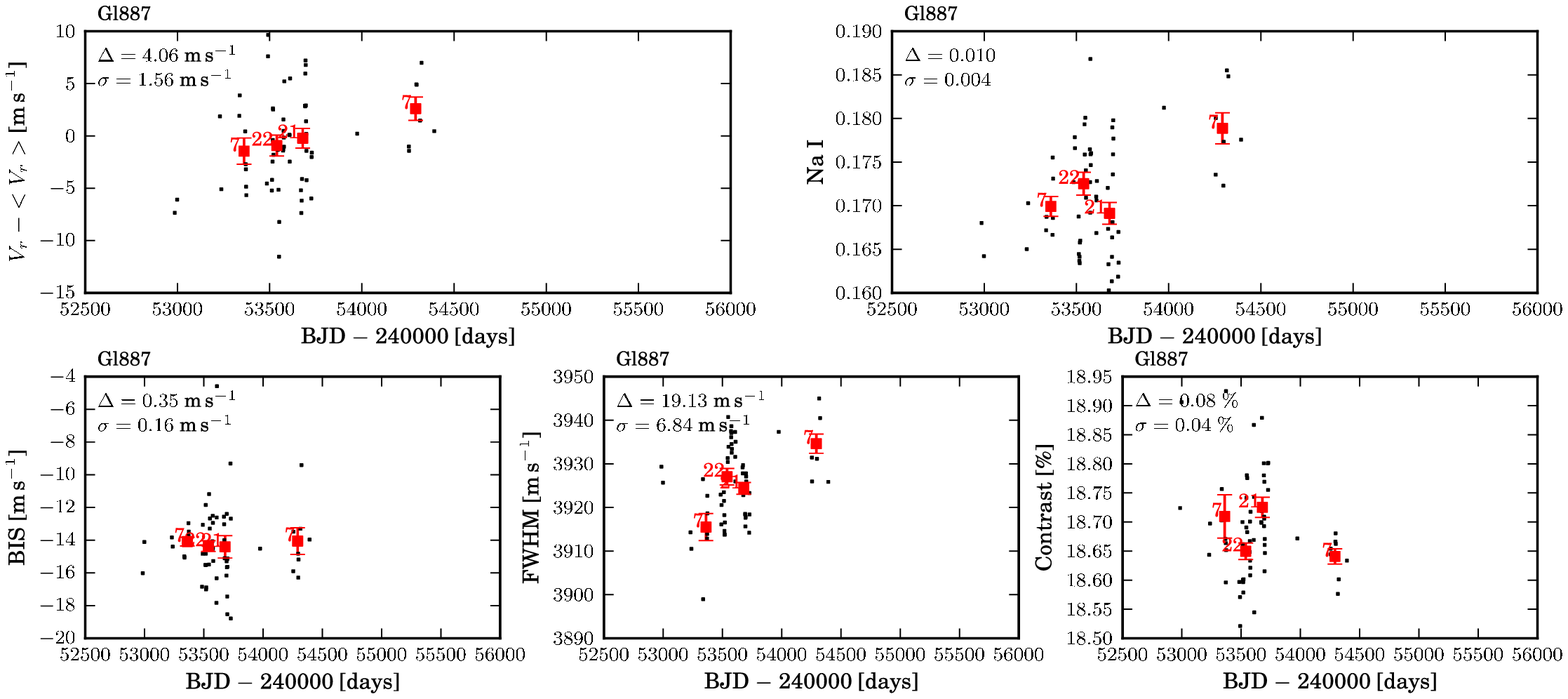}}
\resizebox{\hsize}{!}{\includegraphics{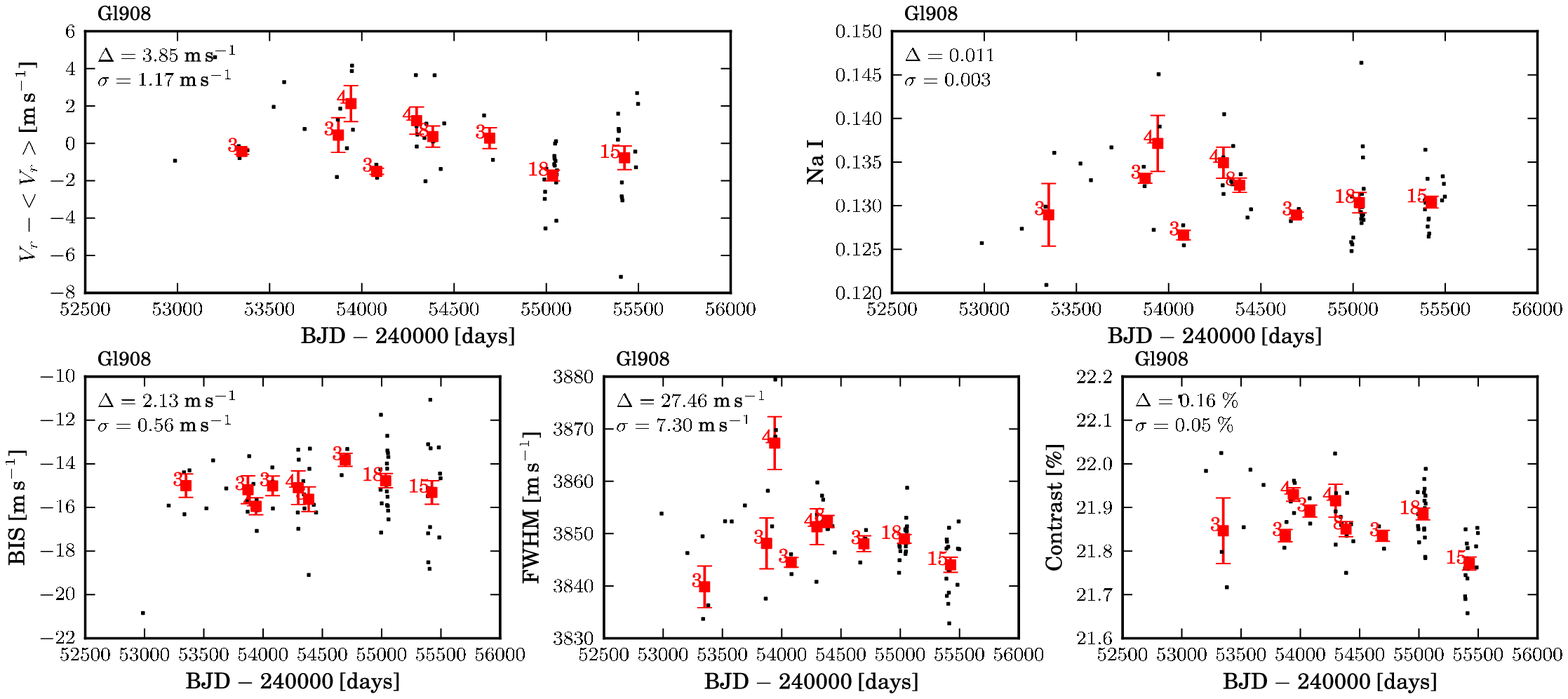}}
\resizebox{\hsize}{!}{\includegraphics{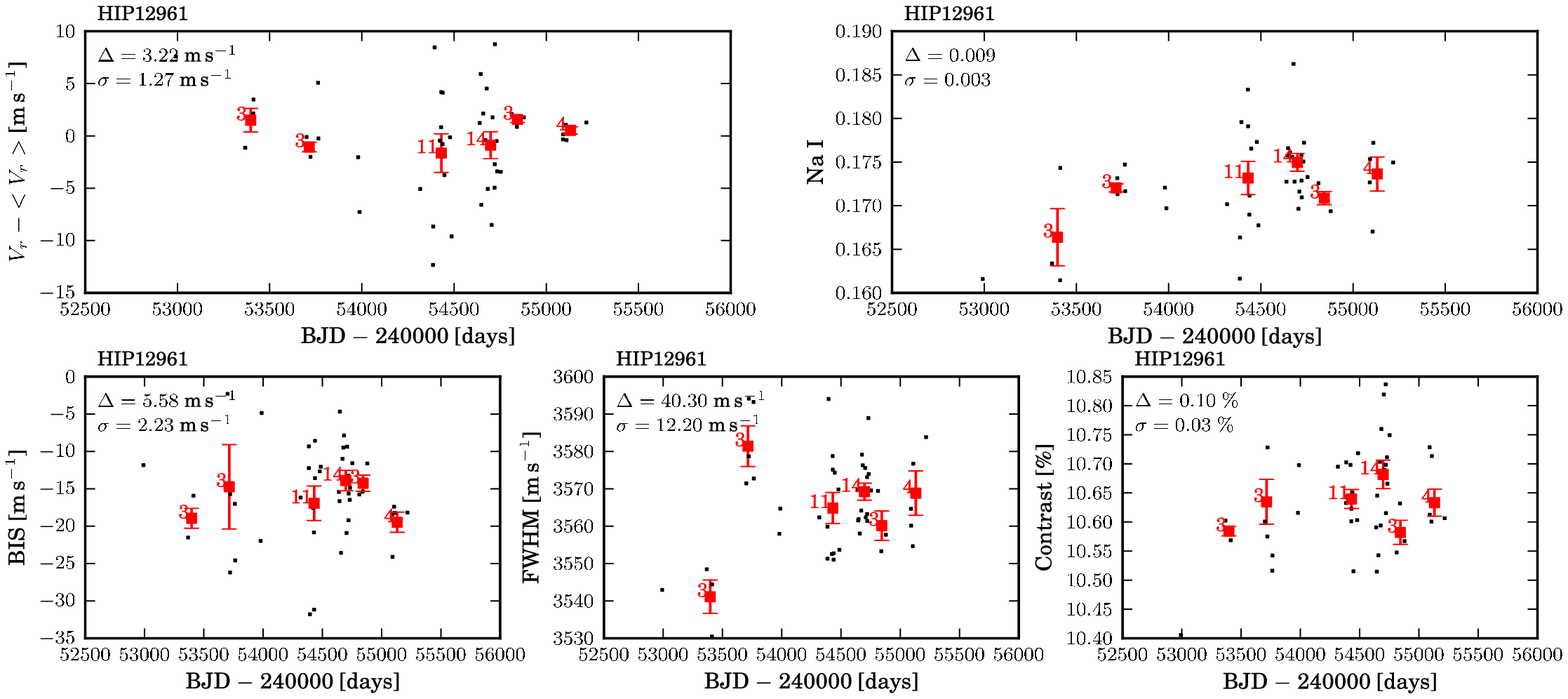}}
\caption{Continued.}
\end{center}
\end{figure*}

\begin{figure*}[tbp]
\ContinuedFloat
\begin{center}
\resizebox{\hsize}{!}{\includegraphics{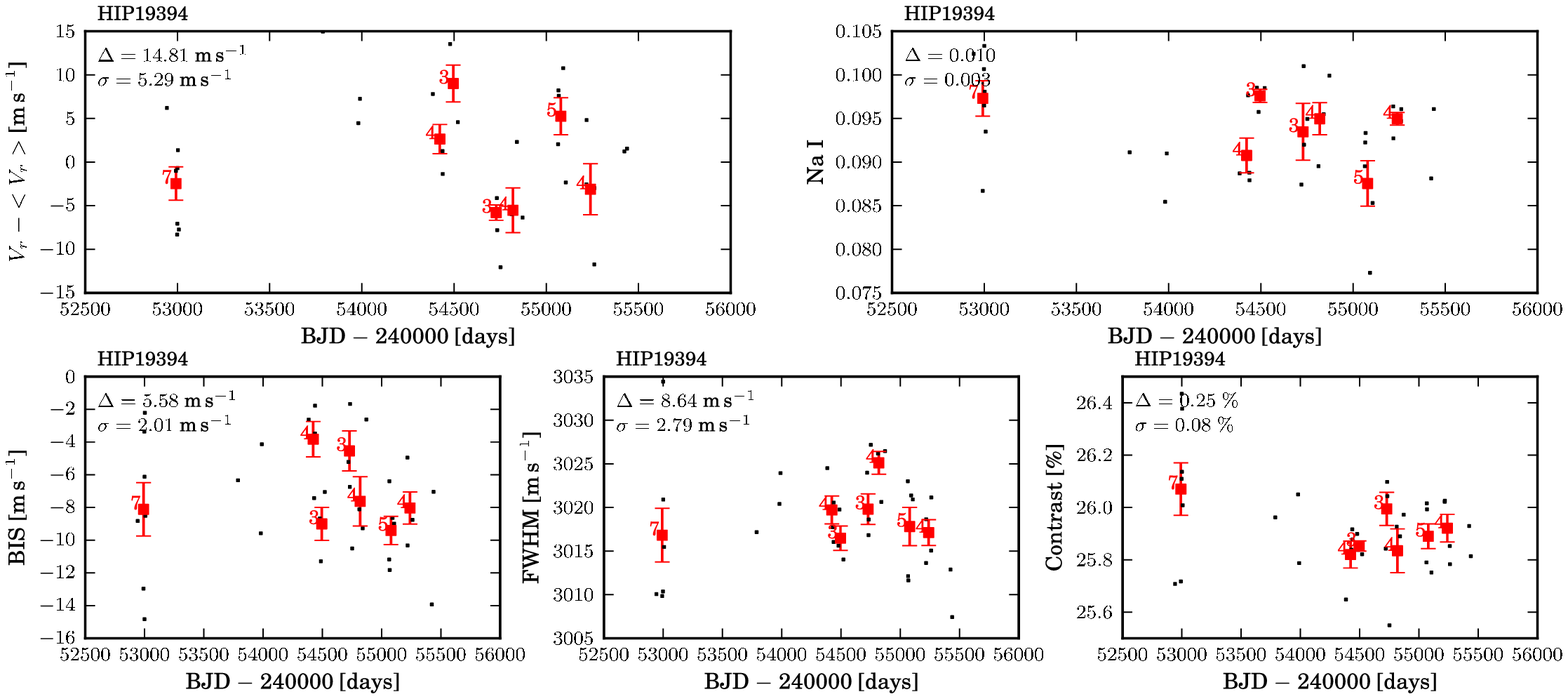}}
\resizebox{\hsize}{!}{\includegraphics{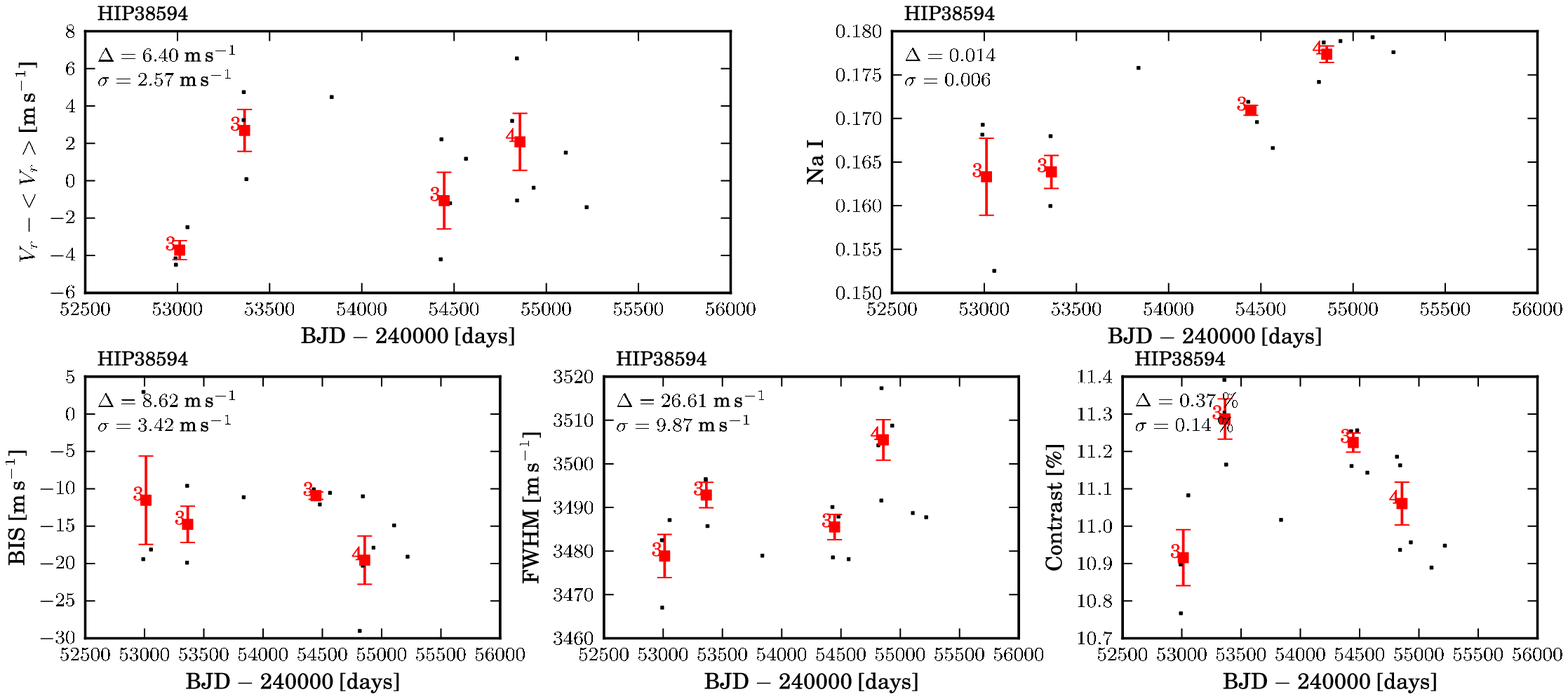}}
\resizebox{\hsize}{!}{\includegraphics{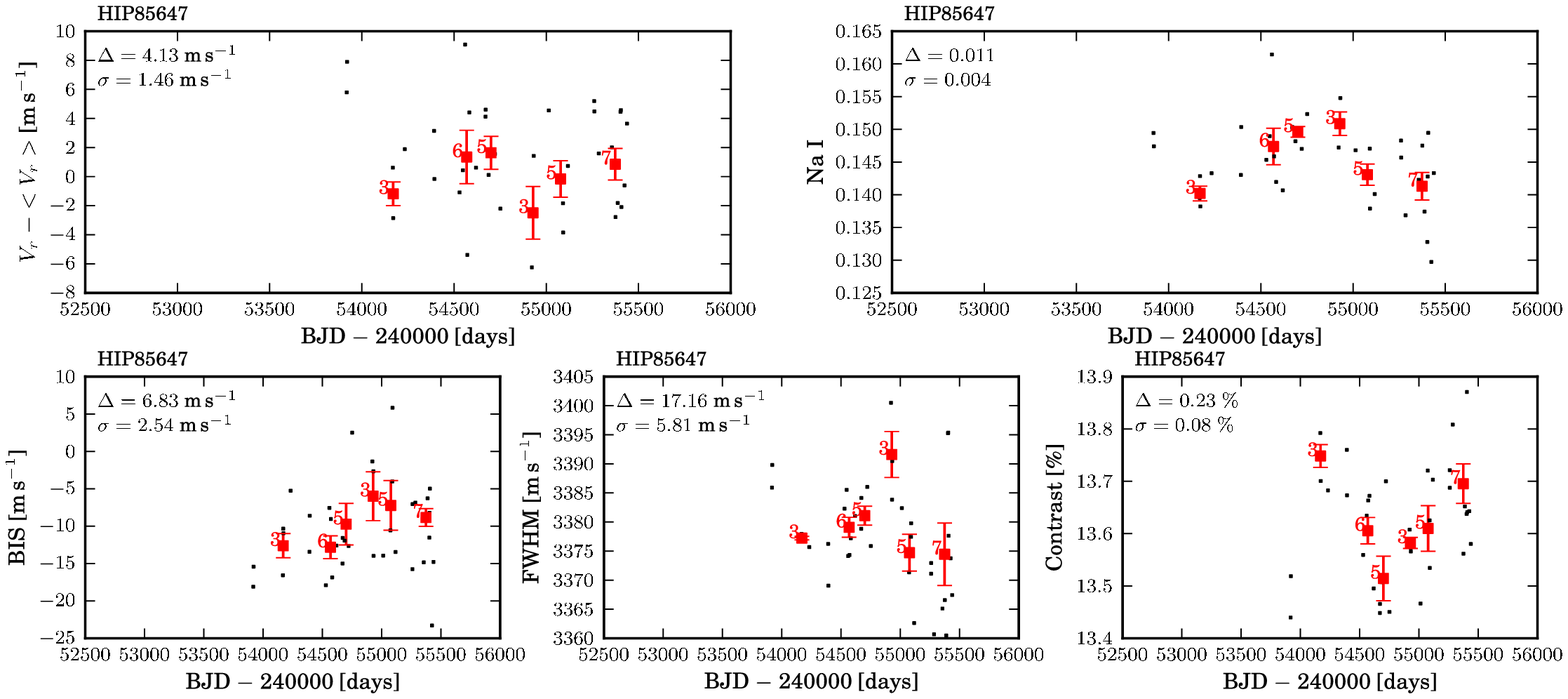}}
\caption{Continued.}
\end{center}
\end{figure*}
}

Figure \ref{all_plots} shows the time-series of radial velocity, activity, and the CCF parameters' BIS, FWHM and contrast for the 27 stars that passed the selection criteria (before the variability F-tests). Small points are nightly averaged data used to calculate the bins, points with errorbars are the binned data (the errors are explained in Sect. \ref{sect:sample}). The peak-to-peak variation ($\Delta$) and rms ($\sigma$) are also shown. The RV time-series have the known planetary companions' signal subtracted.

HIP38594 showed a trend in RV after removal of secular acceleration, with a slope of $-50.83 \pm 0.51$ m\,s$^{-1}$\,yr$^{-1}$.
This corresponds to a variation of $\Delta(V_r)=307.5$ m\,s$^{-1}$ in our timespan and is probably due to a stellar companion.
A keplerian fit gives an orbit for a companion with a minimum period (only a linear trend is observed in the data) of $4941\pm 1516$ days.
Since we are only concerned with low-amplitude and long-term variations, we removed this trend.

Other stars appear to have linear trends in RV, such as Gl\,205 (with $\Delta(V_r) = 9$ m\,s$^{-1}$) and Gl\,680 ($\Delta(V_r) = 15$ m\,s$^{-1}$), but since these trends have lower amplitude variations we decided to keep them for the rest of the study (see Sect. \ref{sect:rv_naI} for more information).

\subsection{Activity and radial-velocity scatter} \label{sect:rvscatter}

\begin{figure}[tbp]
\centering
\resizebox{\hsize}{!}{\includegraphics{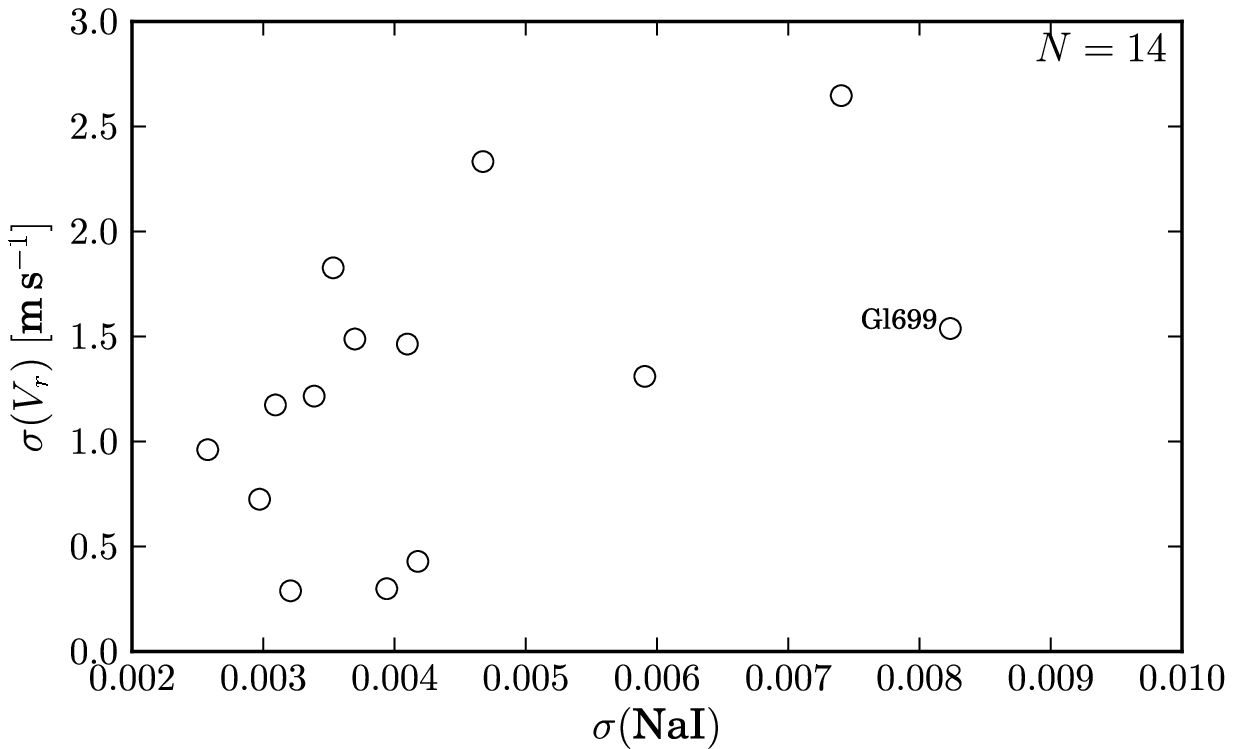}}
\resizebox{\hsize}{!}{\includegraphics{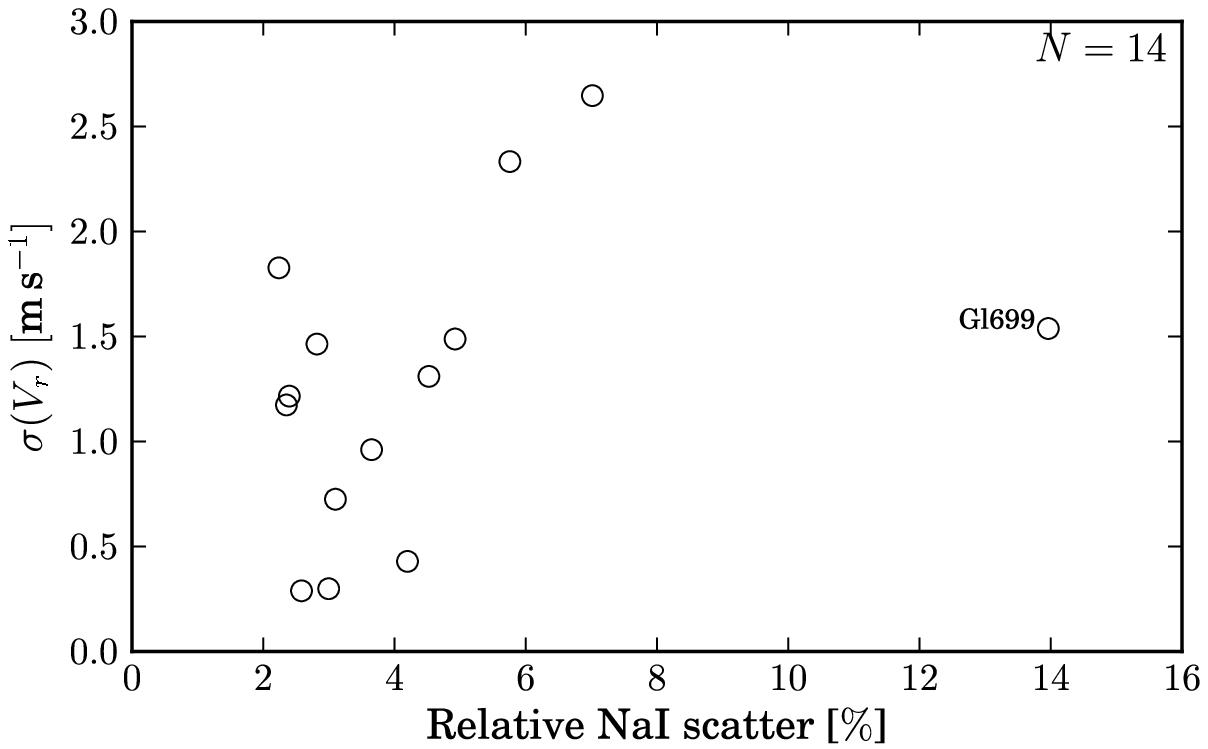}}
\caption{\textit{Top:} Radial-velocity scatter versus \ion{Na}{i} scatter. \textit{Bottom:} Radial-velocity scatter versus relative \ion{Na}{i} scatter.}
\label{sig_rv_sig_naI}
\end{figure}

Table \ref{table:var} shows the radial-velocity rms after the subtraction of the orbits together with the mean and rms of the $\ion{Na}{i}$ index for the binned data of the full sample.
The scatter in radial velocity, $\sigma_e(V_r)$, varies between 0.29 for Gl\,849 and 5.9 m\,s$^{-1}$ for Gl\,680.
The most active star in the sample is Gl\,551 (Proxima Centauri) with an average $\ion{Na}{i}$ index of 0.49, while the most inactive star is Gl\,699 (Barnard's star) with an average $\ion{Na}{i}$ value of 0.059 (but highest relative scatter in \ion{Na}{i} with $\sim$$14 \%$). We should alert, however, that the average \ion{Na}{i} index values depend on stellar color (see Fig. 6 of Paper I) and should not be used to compare activity between stars with different colors without having the index calibrated for the photospheric contribution and without being reported to the bolometric flux.  

The most active star in the sample, Gl\,551, also have the highest absolute scatter in \ion{Na}{i} index (and $12 \%$ relative scatter). But its RV dispersion is not particularly high, having a standard deviation of $\sigma_e(V_r) = 1.5$ m\,s$^{-1}$.

\citet{meunier2010} simulated the effects of spots, plages, and inhibition of convection in the solar radial velocity during a solar activity cycle when the Sun is seen edge-on and observed as a star (integrated flux over the whole disk).
The authors found that the signal induced by the three effects on the solar RV during the cycle had an rms of $\sigma(V_r) = 2.40$ m\,s$^{-1}$, with a peak-to-peak variation of 10.6 m\,s$^{-1}$.

The median RV rms of 1.5 m\,s$^{-1}$ for our sample is lower than the results of \citet{meunier2010} and therefore in agreement with the extrapolation towards M dwarfs of \citet{lovis2011} results which found that later-type stars have lower amplitude RV influenced by long-term activity than earlier-type stars.
It is expected that the stronger contribution to RV from magnetic activity cycles comes from the inhibition of convection \citep{meunier2010}.
Since cell convection structure is strongly depending on the effective temperature of the stars, the impact of their inhibition should be also depending of spectral type.
Thus, it is expected that activity cycles will have different influence on RV for different types of stars.
Furthermore, some of these stars may have undiscovered low-mass companions that could reduce the RV rms even further.

Figure \ref{sig_rv_sig_naI} shows the rms variation of radial velocity against the rms of \ion{Na}{i} index (top panel) and relative variation of \ion{Na}{i} (lower panel).
A relation between these parameters is clearly visible: stars with higher long-term scatter of activity also present higher long-term scatter in radial velocity.
The minimum long-term relative variation we could detect in the \ion{Na}{i} index was around 2\%, with highest scatter reaching around 14\%.
Gl\,699 appears to be an outlier, more evidently when we consider the relative activity values.
Although this star has an average value of $\sigma(V_r)$ when compared to the rest of the sample, it has the highest relative activity variation, almost the two times the value of the star with the second highest relative activity variation.

\subsection{Correlation between RV and activity} \label{sect:rv_naI}

\begin{table*}[ht] 
\caption{Pearson correlation coefficients with respective FAPs.}
\label{table:corr}
\centering
\begin{tabular}{l c c | c c | c c | c c } \\
\hline
\hline
\multicolumn{1}{l}{Star} &
\multicolumn{2}{c}{$V_r$ vs $\ion{Na}{i}$} &
\multicolumn{2}{c}{$\ion{Na}{i}$ vs BIS} &
\multicolumn{2}{c}{$\ion{Na}{i}$ vs FWHM} &
\multicolumn{2}{c}{$\ion{Na}{i}$ vs Contrast} \\
\multicolumn{1}{l}{} &
\multicolumn{1}{c}{$\rho$} &
\multicolumn{1}{c}{FAP} &
\multicolumn{1}{c}{$\rho$} &
\multicolumn{1}{c}{FAP} &
\multicolumn{1}{c}{$\rho$} &
\multicolumn{1}{c}{FAP} &
\multicolumn{1}{c}{$\rho$} &
\multicolumn{1}{c}{FAP} \\
\hline
Gl\,273	&	0.84	&	\textbf{0.022}	&		-0.64	&	0.094		&	0.60	&	0.13			&	-0.89	&	\textbf{0.0074} \\
Gl\,433	&	0.91	&	\textbf{0.0018}	&		-0.70	&	0.11			&	0.70	&	\textbf{0.031}	&	-0.53	&	0.14 \\
Gl\,436	&	0.83	&	\textbf{0.0091}	&		0.28	&	0.29			&	0.82	&	\textbf{0.0074}	&	0.44	&	0.15 \\
Gl\,526	&	0.86	&	0.13			&		0.27	&	0.34			&	0.90	&	0.057		&	-0.76	&	0.13 \\
Gl\,581	&	0.24	&	0.25			&		0.11	&	0.38			&	0.60	&	\textbf{0.019}	&	-0.38	&	0.13 \\
Gl\,588	&	0.82	&	\textbf{0.033}	&		0.03	&	0.49			&	-0.51	&	0.24			&	-0.40	&	0.25 \\
Gl\,667C	&	0.45	&	0.12			&		-0.35	&	0.26			&	-0.27	&	0.22			&	0.37	&	0.17 \\
Gl\,699	&	0.32	&	0.31			&		-0.12	&	0.42			&	0.55	&	0.15			&	-0.79	&	0.070 \\
Gl\,832	&	-0.20	&	0.36			&		-0.41	&	0.30			&	0.57	&	0.19			&	0.08	&	0.46 \\
Gl\,849	&	-0.37	&	0.25			&		-0.38	&	0.26			&	0.69	&	0.097		&	-0.29	&	0.29 \\
Gl\,876	&	0.63	&	0.091		&		0.45	&	0.15			&	0.27	&	0.30			&	-0.15	&	0.39 \\
Gl\,877	&	0.77	&	0.11			&		-0.74	&	0.085		&	0.93	&	\textbf{0.023}	&	-0.45	&	0.21 \\
Gl\,908	&	0.85	&	\textbf{0.0038}	&		-0.61	&	0.051		&	0.82	&	\textbf{0.0075}	&	0.43	&	0.093 \\
HIP85647	&	-0.03	&	0.48			&		0.32	&	0.30			&	0.81	&	\textbf{0.039}	&	-0.87	&	\textbf{0.014} \\
\hline
\end{tabular}
\tablefoot{Only the 14 stars that passed the activity variability tests in Sect. \ref{vartests} are considered. Bold highlights indicate FAP values lower than 0.05 (95\% significance level). FAPs were calculated via bootstraping as in Paper I.}
\end{table*}

Table \ref{table:corr} shows the correlation coefficients of the relation between RV and \ion{Na}{i} with the respective FAPs.
The FAPs were calculated using bootstrapping of the nightly averaged data, followed by a re-binning and a subsequent determination of the correlation coefficient for each of the 10000 permutations (see also Paper I).
The tendency for positive correlations is clear, with an average correlation coefficient of $<\rho> = 0.49$.
Five stars show a significant correlation coefficient (with $\leq 5\%$ false-alarm probability) between activity and radial velocity, which represents 36\% of the sample.
These stars are Gl\,273, Gl\,433, Gl\,436, Gl\,588, and Gl\,908 and their coefficients range between 0.82 and 0.91.
One more star, Gl\,876, has a FAP value lower than 10\%, with a correlation coefficient of 0.63.
If a FAP lower than 10\% is taken as the limit, then we have $\sim$43\% of the variable stars having their RV induced by long-term activity.

\begin{figure}[t]
\centering
\resizebox{\hsize}{!}{\includegraphics{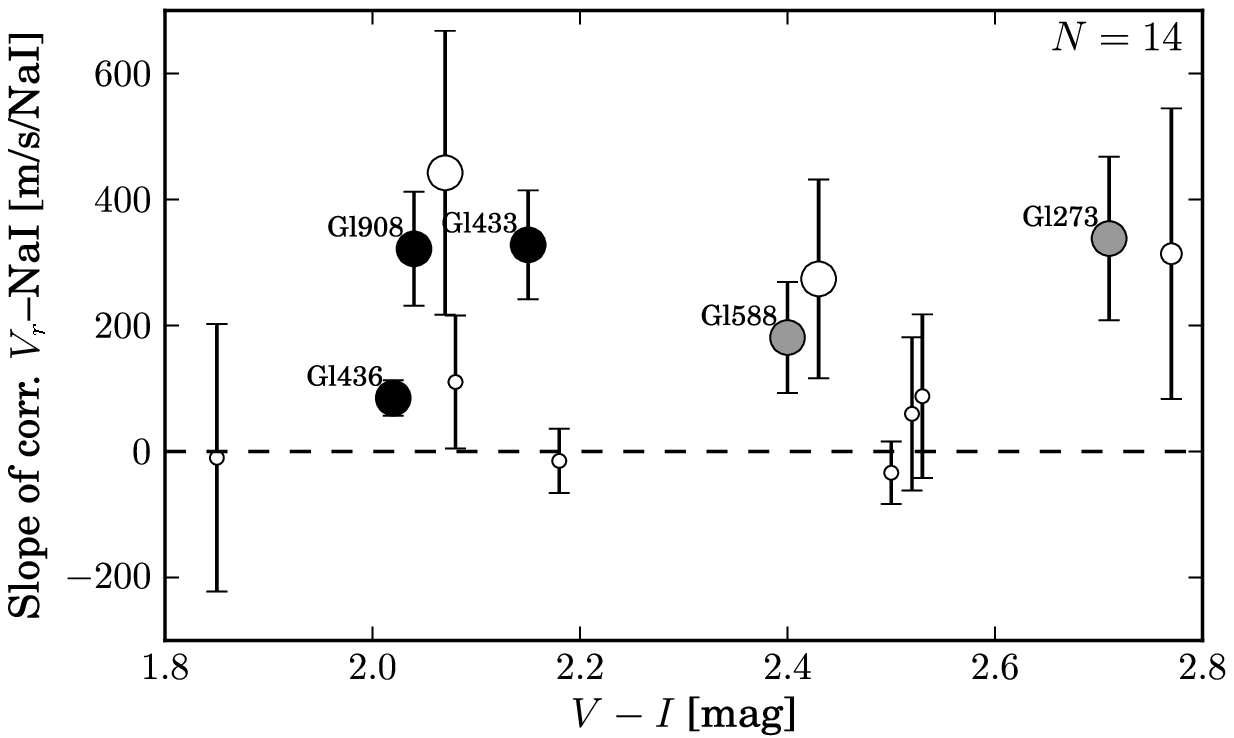}}
\caption{Slope of the correlation between velocity and activity against $(V-I)$ color. Symbol sizes depend on the value of the correlation coefficient between RV and activity: large for $\rho \geq 0.75$, medium for $0.50 \leq \rho < 0.75$, and small for $\rho < 0.50$.
The data points have three colors based on the FAP values: black for FAP $\leq 0.01$, grey for $0.01 <$ FAP $\leq 0.05$, and white for FAP $>  0.1$.
The errorbars are the errors on the slope.}
\label{slope_rv_naI}
\end{figure}

Fig. \ref{slope_rv_naI} shows the slope of the correlation between radial velocity and \ion{Na}{i} index as a function of $(V-I)$ colour.
Symbols are in three sizes dependent on the correlation coefficient values: large for $\rho \geq 0.75$, medium for $0.50 \leq \rho < 0.75$, and small for $\rho < 0.50$.
The data points have three colors based on the FAP values: black for FAP $\leq 0.01$, grey for $0.01 <$ FAP $\leq 0.05$, and white for FAP $>  0.1$.
The errorbars are the errors on the slope.
We can see in the figure that there is a clear tendency for positive slopes of the correlation RV--\ion{Na}{i}.
This implies that a positive change in activity will induce a positive change in the apparent stellar velocity.

All the cases of significant correlation have positive coefficients and the slopes of the correlation range from 85 to 338 m\,s$^{-1}$\,\ion{Na}{i}$^{-1}$.
Contrary to what \citet{lovis2011} found for FGK stars, we found no strong relationship between the slopes and stellar colour (a proxy of $T_{\mathrm{eff}}$) for this sample of early-M dwarfs.
Furthermore, we found no cases of significant anti-correlations as could be expected from the author's study.
It can be that the author's correlation observed in their Fig. 18 (lower panel) reaches a plateau for lower effective temperatures or that the variations induced by long-term activity on the RV of M dwarfs are so low that they are more difficult to observe (some of the observed scatter reaches the same level as the instrument precision).
Another possibility could be that still undetected small planets might interfere with the radial-velocities and therefore ruin any hypothetical correlation.

\subsection{Correlation between activity and the CCF parameters}

\subsubsection{\ion{Na}{i} vs BIS}
In this work we found a marginal tendency for negative correlations between \ion{Na}{i} and BIS, with $<\rho> = -0.18$.
No cases of significative coefficients with FAP $\leq$ 5\%.
Three cases of marginal (5\% $<$ FAP $\leq$ 10\%) anti-correlation, Gl\,273 with $\rho = -0.64$, Gl\,877 with $\rho = -0.74$, and Gl\,908 with $\rho = -0.61$ (21\% of the sample).
Due to the lack of significant cases of correlation with activity, the line bisector seems thus not to be a very good long-term activity indicator for early-M dwarfs.

This is different from the trend found between activity and BIS for early-K stars where the two were found to be positively correlated \citep[e.g.][]{santos2010}.
While the BIS values found here were all negative, \citet{santos2010} measured positive BIS values in all stars.
Although the absolute value of BIS would increase with activity, for the case of negative values found in this study it means that the increase was a negative increase, and therefore anti-correlated with activity.
We can speculate that the difference in the signal of the BIS values might be attributed to, either the bisectores of K and M dwarfs having inverse shapes, or the use of different cross-correlation masks to obtain the CCF line profiles \citep[see e.g.][]{dall2006}.

\subsubsection{\ion{Na}{i} vs FWHM}
We found a tendency for positive correlations between our activity indicator and width of the CCF profile, with $<\rho> = 0.53$ (the strongest average correlation coefficient between parameters).
Similarly, a positive trend was also found for early-K stars \citep[e.g.][]{santos2010}.
And therefore we consider that the qualitative behaviour of FWHM with activity is similar for different spectral types, an effect also shown in \citet{lovis2011}.
Six stars have strong correlations in the range 0.60--0.93 with FAP $\leq$ 1\% (which represent 43\% of our 14-star sample).
These are Gl\,433, Gl\,436, Gl\,581, Gl\,877, Gl\,908, and HIP85647.
Two more stars show marginal correlation, Gl\,526 and Gl\,849.
If we count them, then 57\% of our stars have long-term correlation between activity and FWHM.
This means that the use of FWHM should be useful to detect long-term activity-like variations in M-dwarfs and can be used in complement to other activity proxies.

\subsubsection{\ion{Na}{i} vs contrast}
As was detected by \citet{santos2010} for earlier-type stars, we found a tendency for anti-correlation, $<\rho> = -0.30$ between the activity level and the contrast of the CCF.
But only two cases of significant coefficients, Gl\,273 with $\rho = -0.89$ and HIP85647 with $\rho = -0.87$ were found.
This represents only 14\% of the sample, an indication that the depth of the CCF line is not an optimal measure of the activity level of M dwarfs (at least as measured by the \ion{Na}{i}/\ion{Ca}{ii} lines).
Two other cases of marginal correlation, Gl\,699 with $\rho = -0.79$ and Gl\,908 with $\rho = 0.43$ are present.

\section{Individual cases} \label{indstars}

\subsection{Stars with significant RV--activity correlation}

\begin{figure}[t]
\centering
\resizebox{\hsize}{!}{\includegraphics{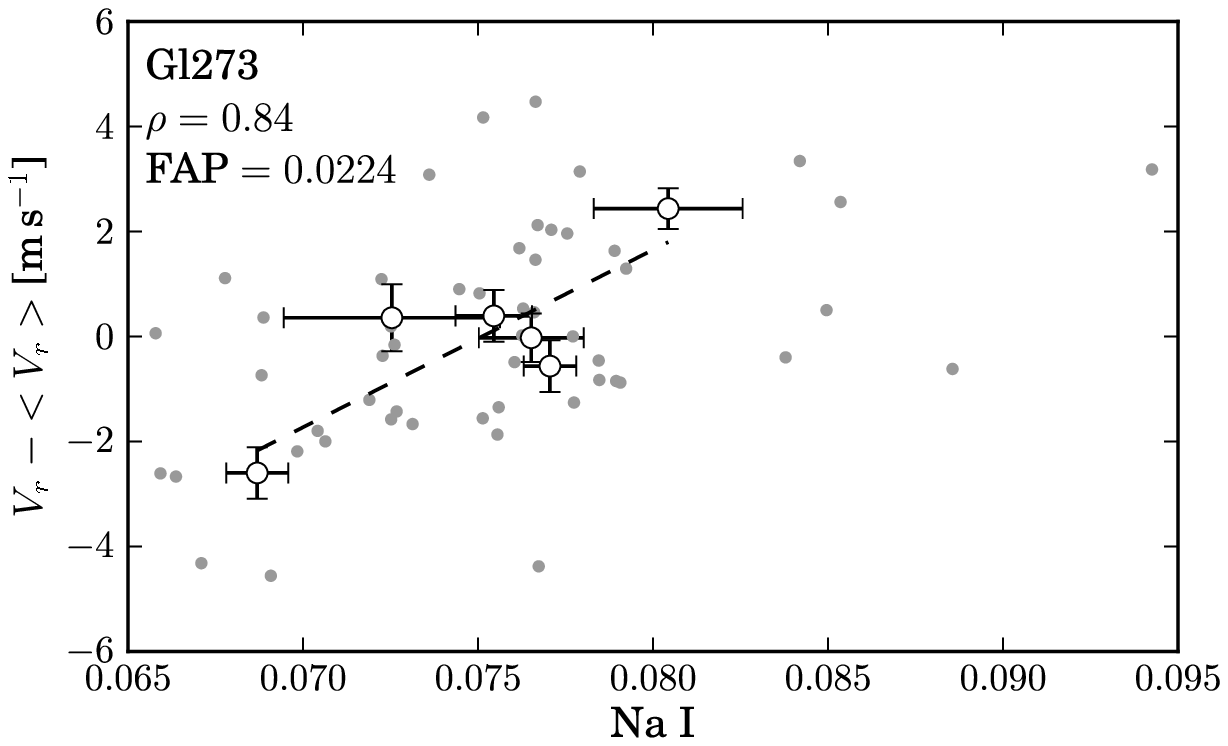}}
\caption{Radial-velocity relation with $\ion{Na}{i}$ index for Gl\,273. Small dots without errorbars are nightly averaged and points with errorbars are averaged over 150 days. The dashed line is the best linear fit. The correlation coefficient $\rho$ and the respective FAP are shown.}
\label{corr_gl273}
\end{figure}

\paragraph{Gl\,273}
This star shows the highest slope of the RV--activity relation of the stars with significant correlation.
Its slope has a value of 338 m\,s$^{-1}$\,\ion{Na}{i}$^{-1}$ with the RV having a peak-to-peak variation of $\Delta = 5$ m\,s$^{-1}$ and a rms of $\sim$1.5 m\,s$^{-1}$.
Its RV--activity correlation coefficient is 0.84 with a FAP of 2\% (Fig. \ref{corr_gl273}).
This star also presents a strong anti-correlation between \ion{Na}{i} and contrast with $\rho = -0.89$ (FAP = 0.0074).
A moderate anti-correlation is also observed between \ion{Na}{i} and BIS ($\rho = -0.64$, FAP = 0.094).
Although we found signs of correlation between activity and RV we were not able to properly fit a sinusoidal to the \ion{Na}{i} time series.

\begin{figure}[t]
\centering
\resizebox{\hsize}{!}{\includegraphics{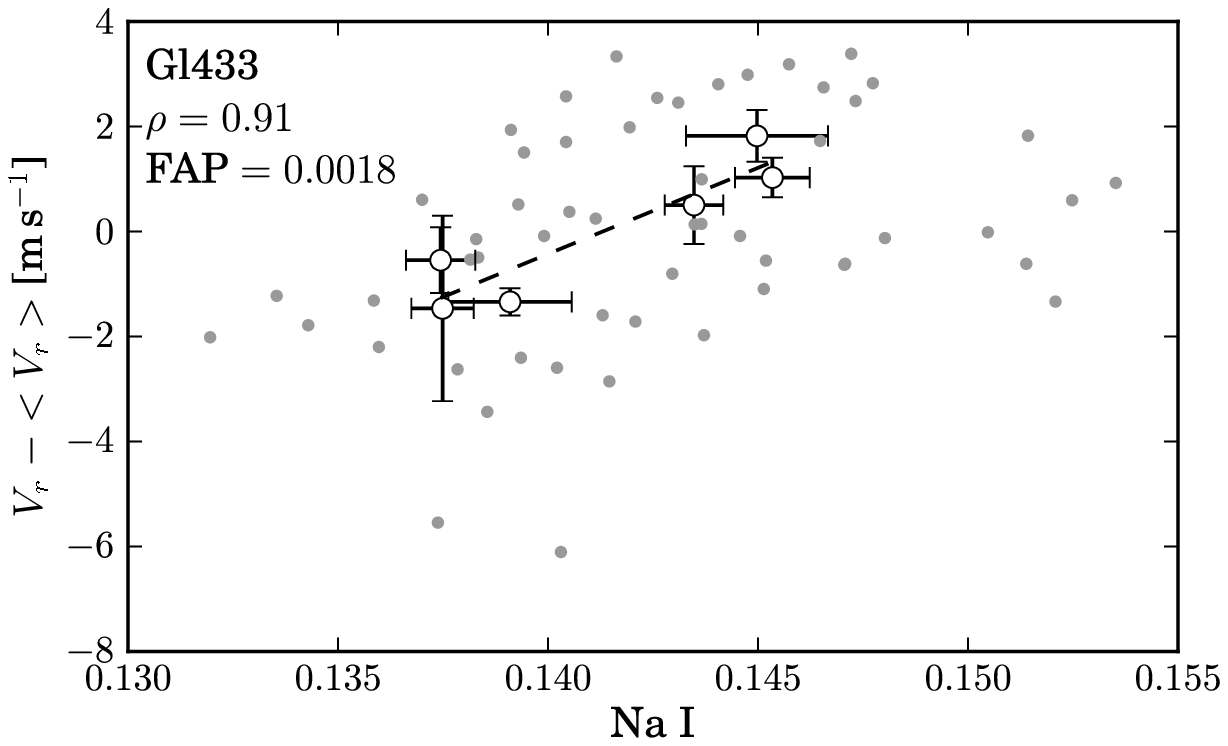}}
\resizebox{\hsize}{!}{\includegraphics{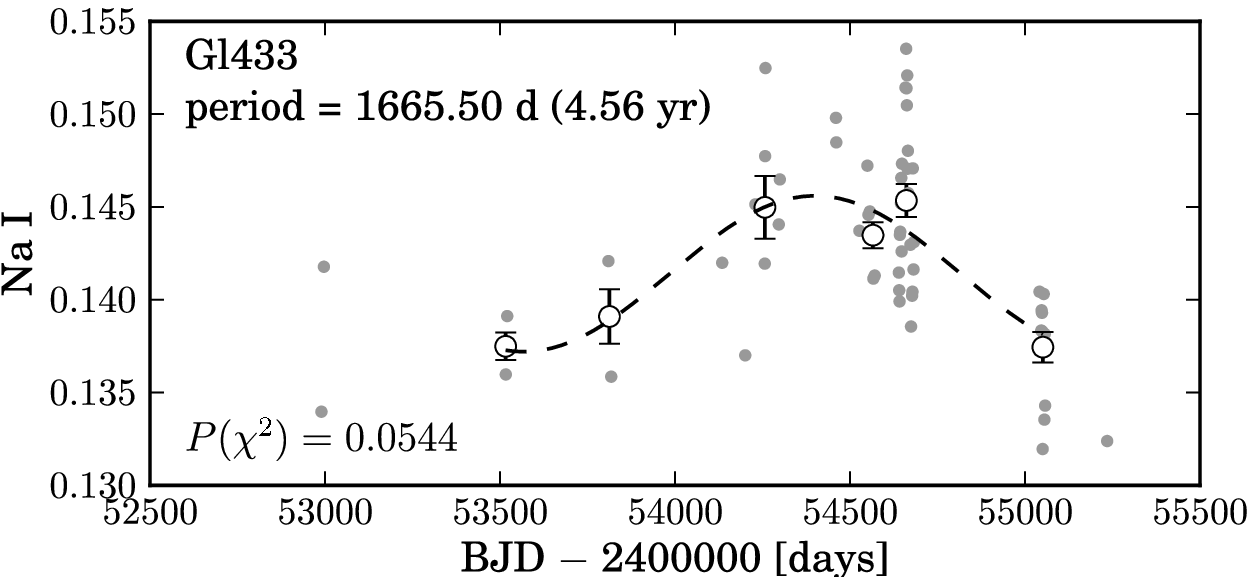}}
\resizebox{\hsize}{!}{\includegraphics{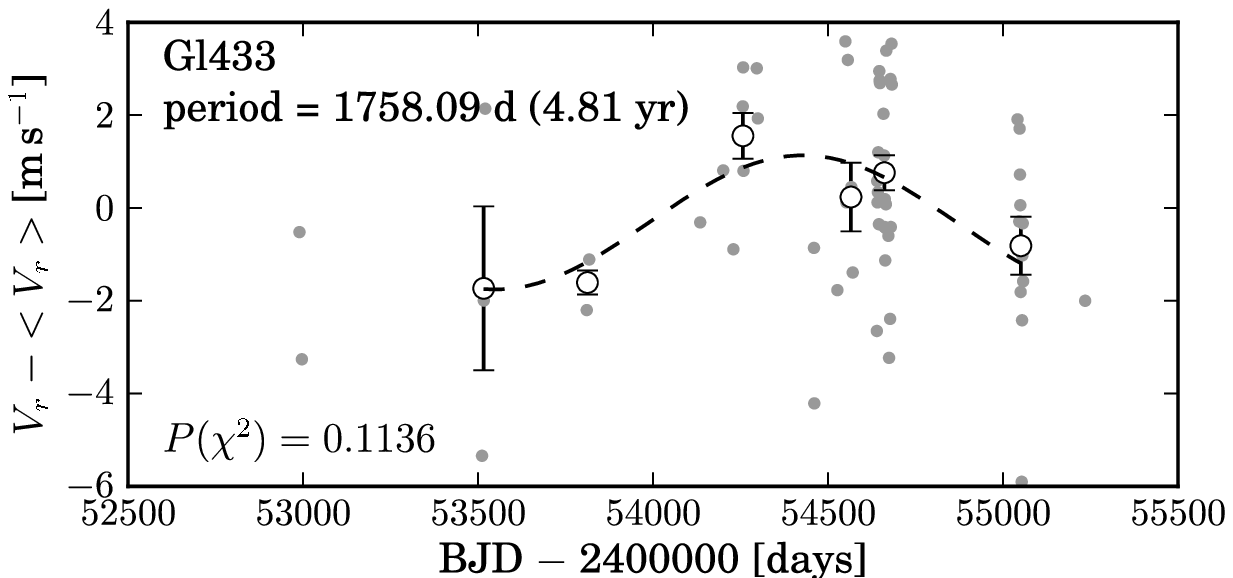}}
\caption{\textit{Top:} Radial-velocity relation with \ion{Na}{i} index for Gl\,433. \textit{Middle:} sinusoidal fit to the \ion{Na}{i} index for Gl\,433. \textit{Bottom:} Sinusoidal fit to RV of Gl\,433. Small dots without errorbars are nightly averaged and points with errorbars are averaged over 150 days. The dashed lines represent the best linear (top) and sinusoidal (middle, bottom) fits to the data.}
\label{gl433}
\end{figure}

\paragraph{Gl\,433}
In Paper I we found this star to present a maximum in activity, typical of cycle-type long-term activity variations.
Figure \ref{gl433} (middle and bottom panels) shows that the RV time-series, after removal of the planetary signal (Delfosse et al. 2012, in prep.), follows a similar pattern as the \ion{Na}{i} index.
The correlation coefficient between the two is $\rho = 0.91$ with a FAP of 0.0018 (Fig. \ref{gl433}, top panel).
The \ion{Na}{i} index also shows correlations with CCF parameters of $-0.70$ and 0.70 for BIS and FWHM respectively.
We can therefore confirm that the activity cycle maximum is inducing a maximum in RV.
The slope of the RV--\ion{Na}{i} correlation is 328 m\,s$^{-1}$\,\ion{Na}{i}$^{-1}$ and we measured an overall variation in RV of 3.29 m\,s$^{-1}$ (with $\sigma_e (V_r) = 1.22$ m\,s$^{-1}$).
The amplitude and period of the cycle in both \ion{Na}{i} or RV scales cannot be inferred because we do not have an entire cycle period in our data.
However, a minimum limit for the amplitude and period can be calculated by fitting a sinusoid to the activity time-series.
We determined that the minimum activity cycle period is 1665 days with a minimum amplitude of 0.004 in \ion{Na}{i} (Fig. \ref{gl433}, middle panel).
A similar minimum period was also found for the radial velocity signal with a value of 1758 days and an amplitude of 1.45 m\,s$^{-1}$ (Fig. \ref{gl433}, bottom panel).
This star represents therefore a good example of an RV signal induced by an activity cycle.

\begin{figure}[t]
\centering
\resizebox{\hsize}{!}{\includegraphics{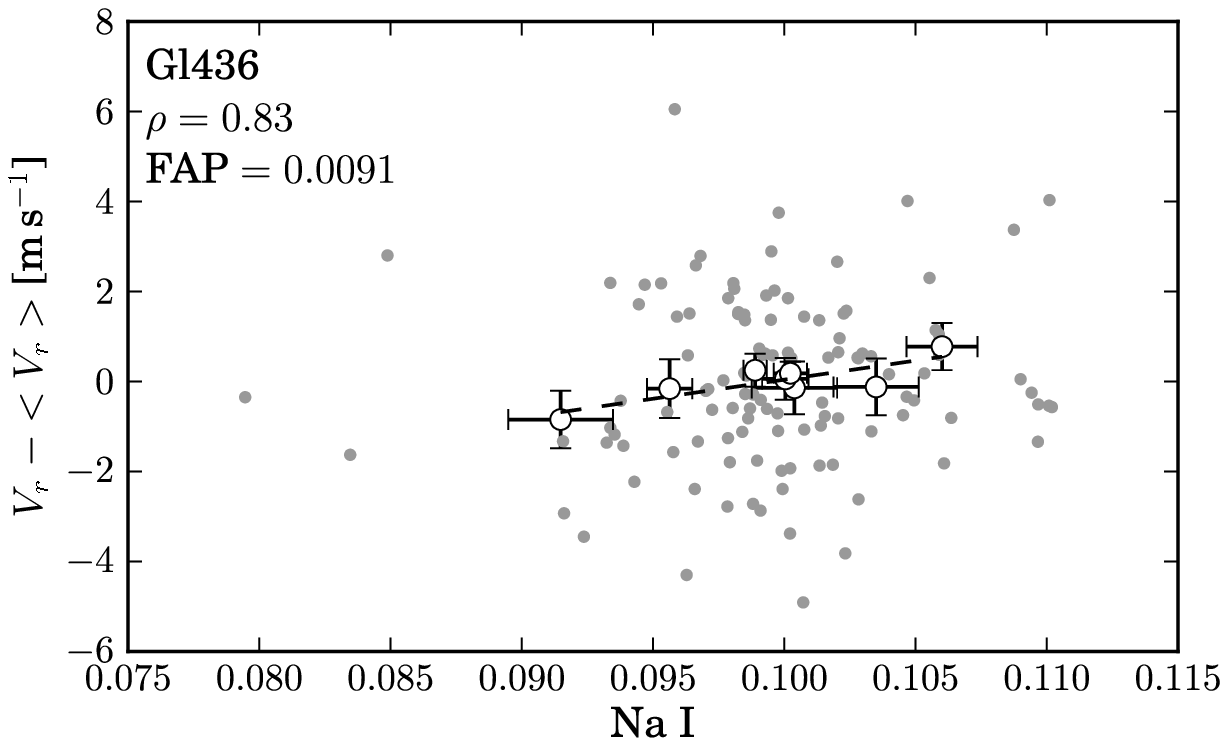}}
\caption{Radial-velocity relation with $\ion{Na}{i}$ index for Gl\,436. Small dots without errorbars are nightly averaged and points with errorbars are averaged over 150 days. The dashed line is the best linear fit. The correlation coefficient $\rho$ and the respective FAP are shown.}
\label{corr_gl436}
\end{figure}

\paragraph{Gl\,436}
This star has the smallest slope of the RV--activity correlation with a value of 85 m\,s$^{-1}$\,\ion{Na}{i}$^{-1}$.
From Fig. \ref{corr_gl436} we can observe a large scatter in the nightly averaged data.
\citet{ballard2010} reported on short-term noise in photometry that they attributed to stellar spots.
Furthermore, \citet{knutson2011} detected evidence of occulted spots during the transits of Gl\,436\,b \citep{butler2004}.
It is probable that short-term activity variability is producing the large scatter in \ion{Na}{i} and contributing to a reduction of the slope's value.
Nevertheless, the correlation coefficient of the RV--activity relation is $\rho = 0.83$ with a FAP of 0.9\%, providing clues for the influence of long-term activity on the observed RV of the star.
A strong correlation coefficient between \ion{Na}{i} and FWHM of 0.82 (FAP = 0.7\%) was also found.
However, we tried to fit a sinusoidal signal to both the RV and \ion{Na}{i} time series but obtained no significant results.
The activity time series in Fig. \ref{all_plots} shows a decreasing trend in activity that can be due to a long-period magnetic cycle, but more data will be needed to firmly prove this.

\begin{figure}[t]
\centering
\resizebox{\hsize}{!}{\includegraphics{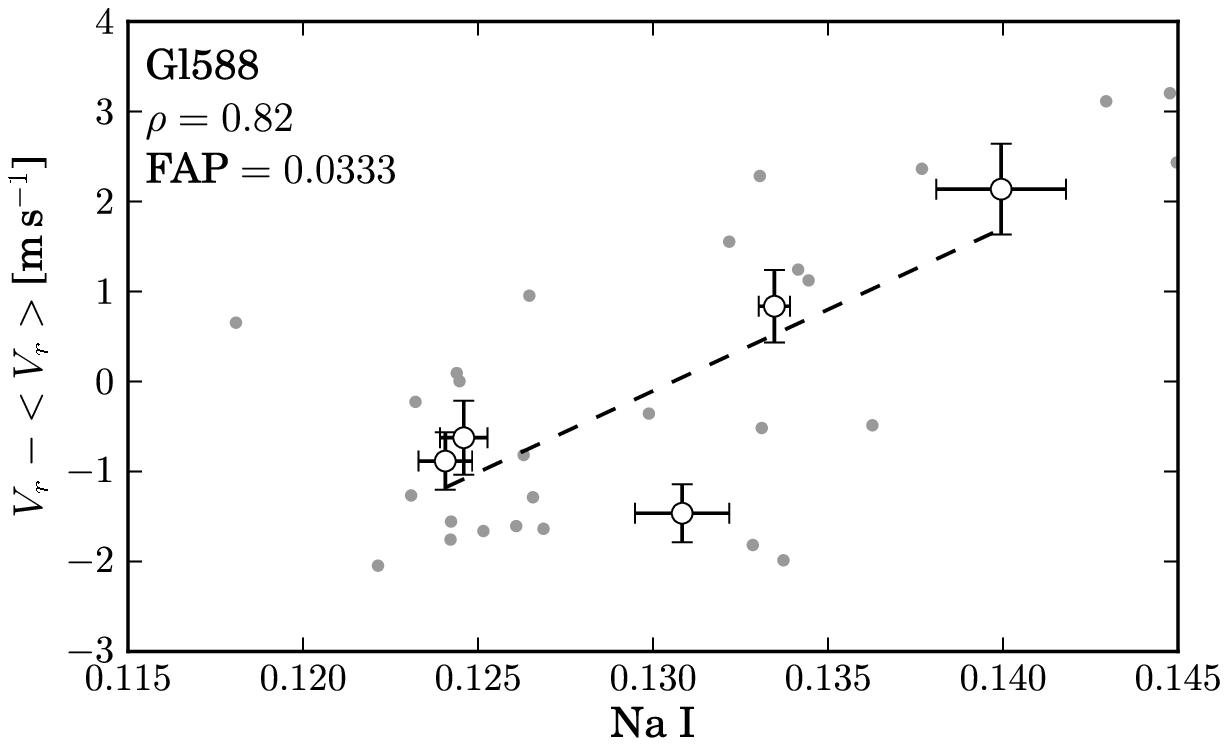}}
\caption{Radial-velocity relation with $\ion{Na}{i}$ index for Gl\,588. Small dots without errorbars are nightly averaged and points with errorbars are averaged over 150 days. The dashed line is the best linear fit. The correlation coefficient $\rho$ and the respective FAP are shown.}
\label{corr_gl588}
\end{figure}

\paragraph{Gl\,588}
With a correlation coefficient of 0.82 (FAP = 3.3\%) this star also presents a strong long-term relationship between velocity and activity (Fig. \ref{corr_gl588}).
This correlation has a slope of 181 m\,s$^{-1}$\,\ion{Na}{i}$^{-1}$.
The RV time series (Fig. \ref{all_plots}) shows a periodic-like signal with a peak-to-peak variation of 3.6 m\,s$^{-1}$ similar to the \ion{Na}{i} time series apart from a point at $BJD \sim 2453560$ d.
This is a point which includes only four nights that might be influenced by short-term activity variability.
It is therefore not possible to conclude about the periodicity of this stars' activity cycle.

\begin{figure}[t]
\centering
\resizebox{\hsize}{!}{\includegraphics{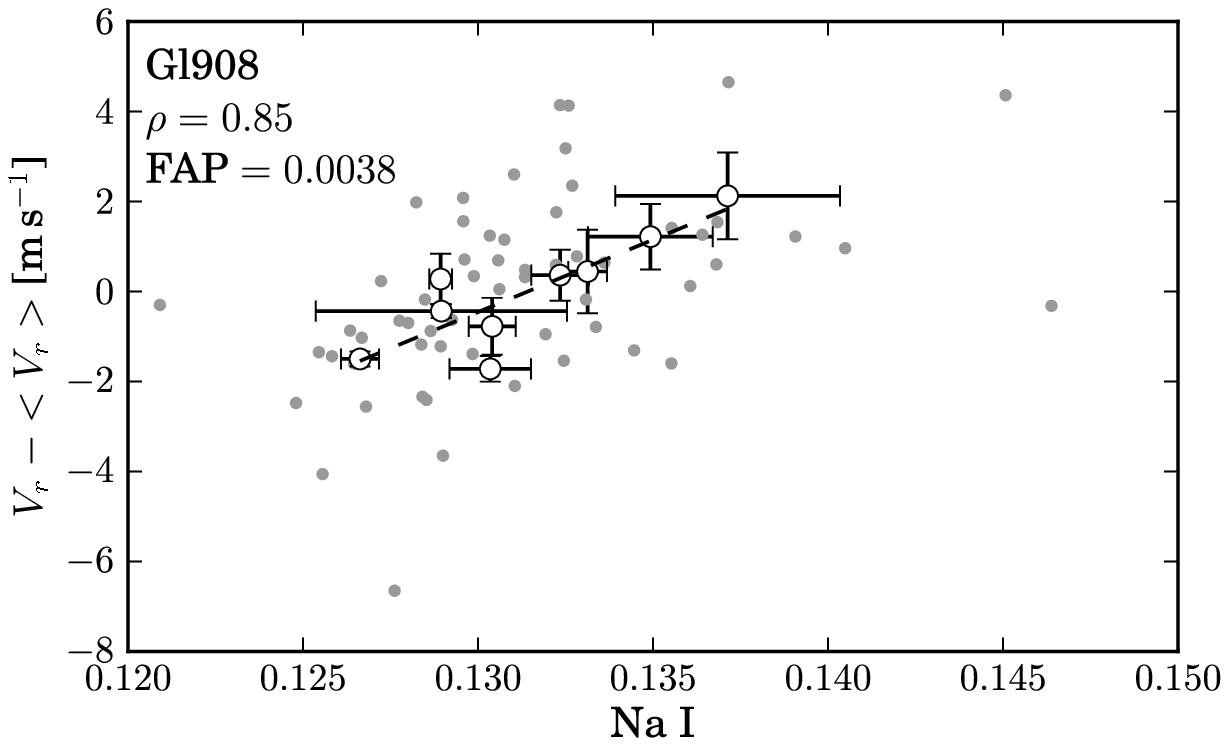}}
\caption{Radial-velocity relation with $\ion{Na}{i}$ index for Gl\,908. Small dots without errorbars are nightly averaged and points with errorbars are averaged over 150 days. The dashed line is the best linear fit. The correlation coefficient $\rho$ and the respective FAP are shown.}
\label{corr_gl908}
\end{figure}

\paragraph{Gl\,908}
This is another good example of long-term correlation between RV and activity.
The correlation coefficient has a value of 0.85 (FAP = 0.4\%) and the trend can be observed both in the binned and in the nightly averaged data (Fig. \ref{corr_gl908}).
The slope of the correlation is 322 m\,s$^{-1}$\,\ion{Na}{i}$^{-1}$ and we measured an RV peak-to-peak variation of 3.85 m\,s$^{-1}$.
A strong correlation between \ion{Na}{i} and FWHM with $\rho = 0.82$ (FAP = 0.75\%) was also detected.
Although we have strong evidence for RV induced by long-term activity, we could not fit a sinusoidal signal to both time series with significant $p(\chi^2)$.
This star has probably a magnetic "cycle" with no determined period, i.e., a star with a long-term aperiodic magnetic activity variability \citep[these types of variations were classified as \textit{var} (variable) by][]{baliunas1995}.

\subsection{Other interesting cases}

\paragraph{Gl\,176}
This star have what appears to be a long-term periodic signal with $P \sim$2043 d in RV (see time-series plot Fig. \ref{all_plots}).
The same trend is not observed in the \ion{Na}{i} index and it did not passed the variability tests.
This stars has a confirmed planetary companion, Gl\,176\,b with $P = 8.7$ d, and a rotationally modulated activity signal with a period of $P = 39$ d \citep{forveille2009}.
But none of these signals can explain the $\sim$2000 d variation.
It can be due to a long-period yet undiscovered planet.

\begin{figure}[tbp]
\centering
\resizebox{\hsize}{!}{\includegraphics{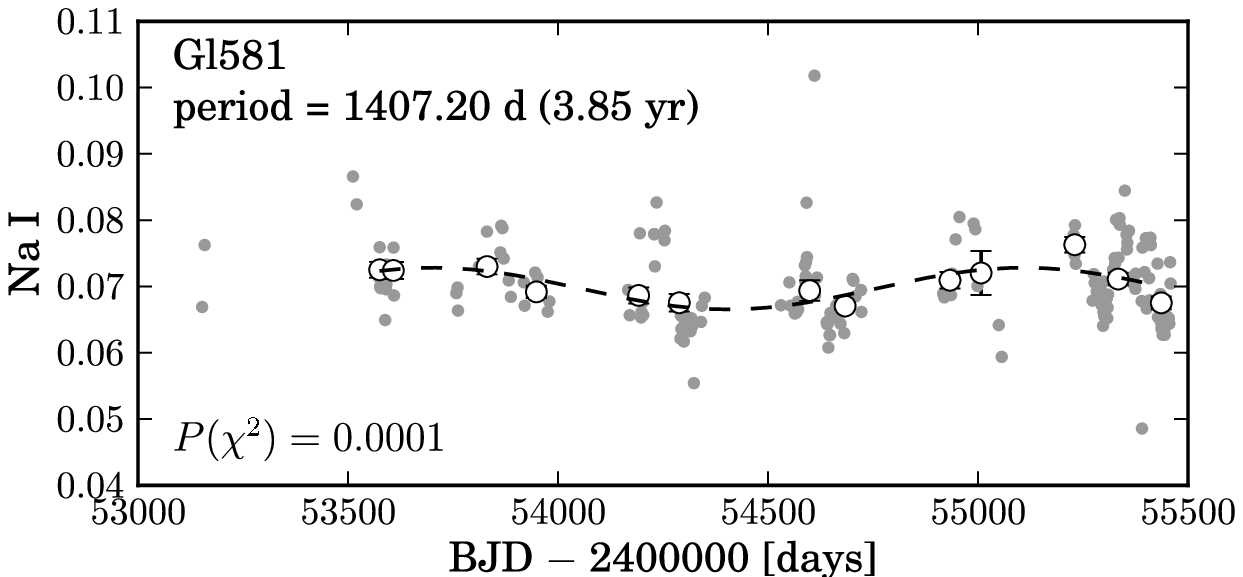}}
\caption{Sinusoidal fit to the activity time series of Gl\,581.}
\label{fig:gl581_cycle}
\end{figure}

\paragraph{Gl\,581}
This is the star which has the best probability of having the activity data fitted by a periodic sinusoidal signal instead of a linear trend.
We obtained a signal with a period of $P=3.85$ years ($P(\chi^2) = 0.01\%$) and our timespan for this star is long enough to cover more than one activity cycle period (Fig. \ref{fig:gl581_cycle}).
Furthermore, Gl\,581 also passed the variability tests.
To date, four planets are known to orbit the star with two more planets that were discarded \citep{bonfils2005,udry2007,mayor2009,forveille2011a}.
 We removed the signal of the four confirmed planets which resulted in a long-term RV rms of 0.96 m\,s$^{-1}$ which is of the order of the instrument precision.
 No correlation was found between long-term velocity and activity but the \ion{Na}{i} index is correlated with FWHM ($\rho = 0.60$, FAP = 1.9\%).
 Because of the large number of detected planets, it is sensible to recognise that their orbits might not have been properly subtracted from the data.
 We will therefore not take any conclusions about the possibility of the activity cycle being influencing the RV signal of this star.  

\begin{figure}[tbp]
\centering
\resizebox{\hsize}{!}{\includegraphics{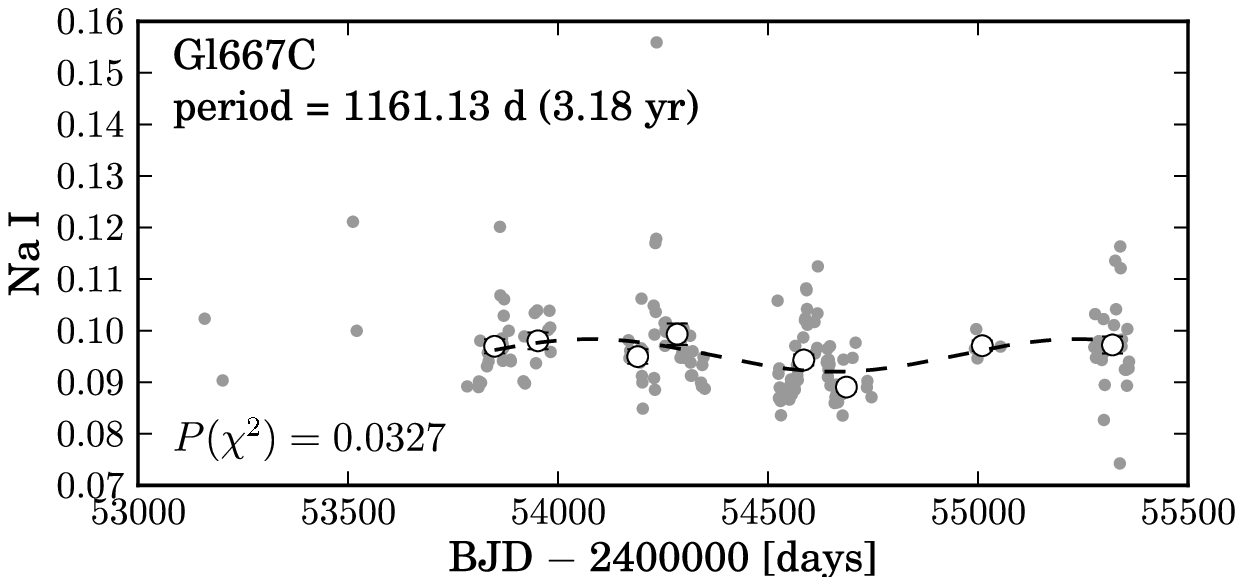}}
\caption{Sinusoidal fit to the activity time series of Gl\,667\,C.}
\label{fig:gl667c_cycle}
\end{figure}

\paragraph{Gl\,667C}
One planet is known to orbit this star (Delfosse et al. 2012, in prep.).
It is also a member of a triple system and orbits the A + B binary system.
This system introduces a trend in RV that was subtracted together with the planetary companion signal.
The \ion{Na}{i} activity index passed the variability F-test and a sinusoidal function with a period of $P = 3.18$ years is well fitted to the activity time series (with $P(\chi^2) = 3.3\%$, Fig. \ref{fig:gl667c_cycle}).
However, the RV signal is only marginally correlated with activity, having a correlation coefficient of $\rho = 0.45$ with a FAP of 12\%.
No correlations between the other parameters are detected.

\paragraph{Gl\,699}
\citet{kurster2003} found an anti-correlation between RV and the H$\alpha$ index for the Bernard's star with a coefficient of $\rho = -0.50$.
Using a longer timespan for the same star, \citet{zechmeister2009} detected a similar correlation with $\rho = -0.42$.
We found only a marginal positive correlation with the \ion{Na}{i} index ($\rho = 0.32$, FAP = 0.32).
The \ion{Na}{i} index of this star is moderately correlated with both FWHM ($\rho = 0.55$, FAP = 0.15) and contrast ($\rho = -0.79$, FAP = 0.070).

\begin{figure}[tbp]
\centering
\resizebox{\hsize}{!}{\includegraphics{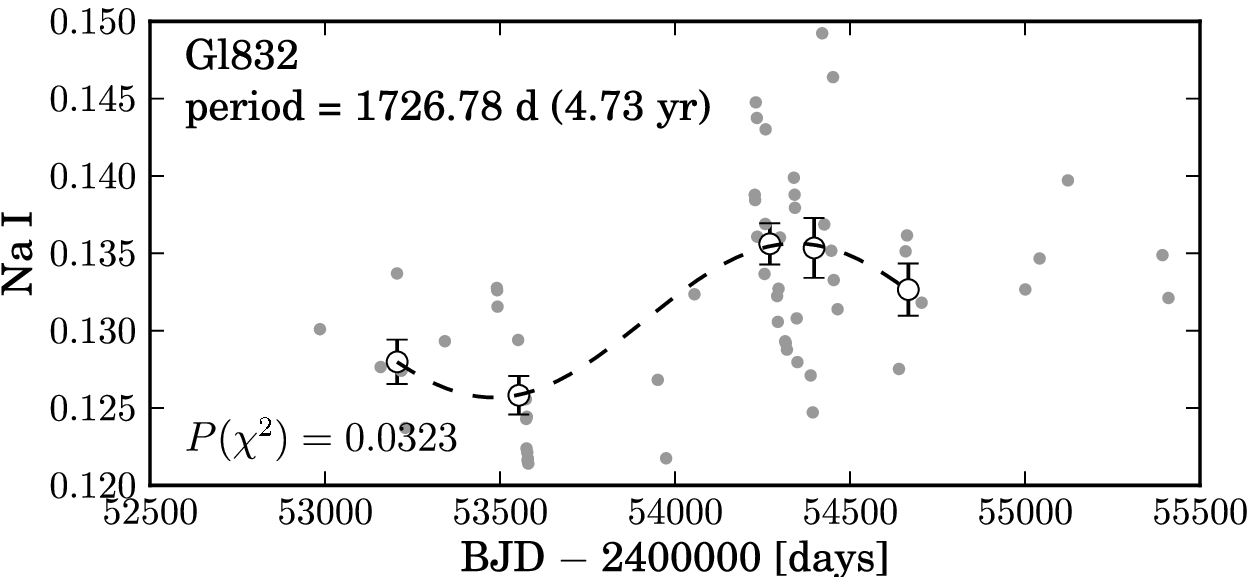}}
\caption{Sinusoidal fit to the activity time series of Gl\,832.}
\label{fig:gl832_cycle}
\end{figure}

\paragraph{Gl\,832}
This star passed the activity variability tests but we found no correlation with radial velocity after the removal of the planetary companion \citep{bailey2009}.
None of the other parameters are correlated apart from a marginal correlation between activity and FWHM of $\rho = 0.57$ but with a high FAP of 19\%.
The activity time series is well fitted by a sinusoidal function with $P(\chi^2) = 3.2\%$ and a minimum period of 4.73 years (Fig. \ref{fig:gl832_cycle}).

\begin{figure}[tbp]
\centering
\resizebox{\hsize}{!}{\includegraphics{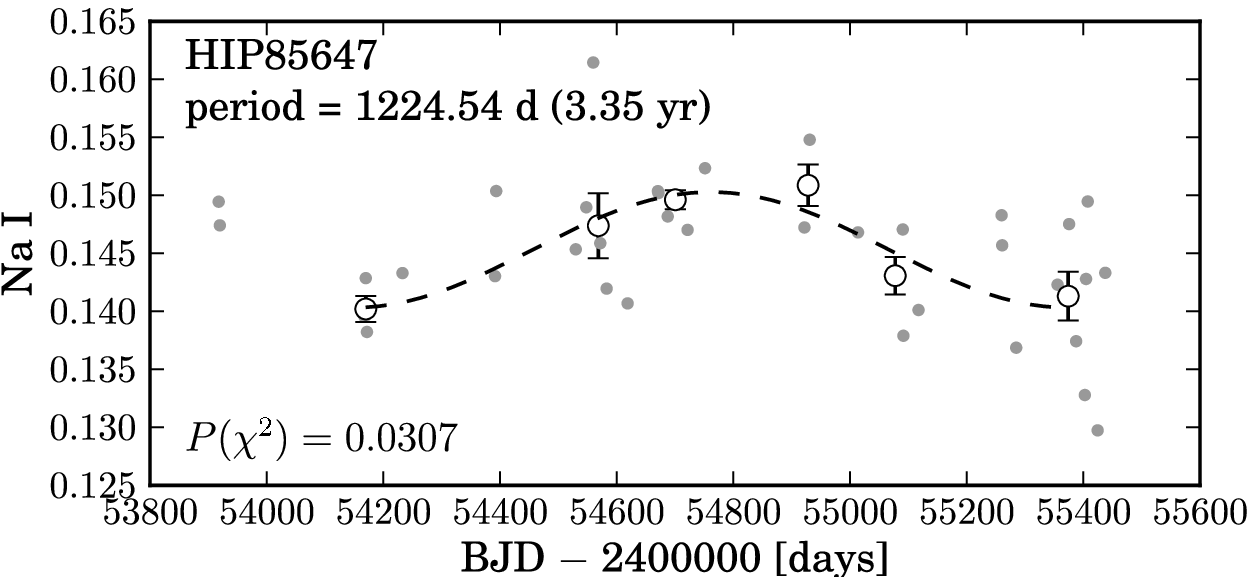}}
\caption{Sinusoidal fit to the activity time series of HIP85647.}
\label{fig:hip85647_cycle}
\end{figure}

\paragraph{HIP85647}
After removal of the signal of the planetary companion \citep{forveille2011b} of this star and a linear trend due to a stellar companion, we found no correlation between RV and activity.
There are strong correlations between \ion{Na}{i} and the CCF parameters FWHM and contrast with $\rho = 0.81$ (FAP = 3.9\%) and $\rho = -0.87$ (FAP = 1.4\%), respectively.
We successfully fitted a sinusoidal signal to the activity time series and obtained a period of 2.74 years (with $P(\chi^2) = 3.4\%$, Fig. \ref{fig:hip85647_cycle}).

\section{Conclusions} \label{conclusions}
We used a sample of 27 M0 to M5.5 dwarfs to study the relationship between long-term activity, radial velocity, and parameters of the cross-correlation function given by the HARPS pipeline.
As the indicator for activity level we used the \ion{Na}{i} index as suggested in Paper I.
We binned the data to 150-day bins to average out high-frequency noise and removed any RV signals induced by known stellar or planetary companions.

A selection of stars having long-term activity variability was carried out by using F-tests.
This resulted in a subsample of 14 variable stars, which means that around half of the stars in our sample have significant long-term activity variations.

The activity time series of some stars were fitted by sinusoidal functions to infer the activity cycle's period or minimum period.
Five stars could be statistically well fitted by sinusoids and the inferred periods varied between 2.8 and 4.7 years.
We should note, however, that our data covered close to one period or less of the fitted signals and therefore the cyclic nature of the signals cannot be fully established.
Even if these stars have cyclic activity variability the periods obtained by the sinusoidal fitting should be regarded as minimum periods since the timespan is not long enough to cover two periods.

We obtained long-term velocity rms in the range 0.30 to 5.9 m\,s$^{-1}$ and a relative scatter in \ion{Na}{i} index for the 14-star variable subsample was in the range $\sim$2--14\%.
These two parameters, $\sigma(V_r)$ and $\sigma(\ion{Na}{i})$, appear to be correlated.
The median RV rms of 1.5 m\,s$^{-1}$ that can possibly be induced by activity is lower than the one obtained for the case of the Sun by \citet{meunier2010} as predicted by extrapolation of  the trend in Fig. 18 (lower panel) of \citet{lovis2011} for late-type stars.

We then searched for correlations between RV, activity and the CCF parameters BIS, FWHM, and contrast.
The main general results we obtained can be summarised as:

\begin{itemize}

\item We found a general tendency for positive correlations between long-term activity and radial velocity variations. No relation between the slope of the correlation and stellar color was found.

\item Five out of 14 stars with long-term variability have significant correlation between activity and radial velocity.
This amounts to 36\% of our subsample.
These stars are Gl\,273, Gl\,433, Gl\,436, Gl\,588, and Gl\,908.
The maximum peak-to-peak RV variation we obtained for stars with significant correlation between long-term activity and RV was $\sim$5 m\,s$^{-1}$.
Only Gl\,433 had its activity well fitted by a sinusoidal signal even though the signal spans less than a full period.
We determined a minimum period of 4.6 years for the activity cycle of this star, and similar minimum period of 4.8 years was found for the RV signal.
The other stars could have aperiodic activity cycles but more data is needed to confirm this hypothesis.

\item
Although we found a general tendency for BIS to be anti-correlated with activity, those correlations were not statistically significant.
Only 21\% of the stars with long-term variability showed marginal correlations between activity and BIS.
While we found a general trend for anti-correlations between activity and BIS for M dwarfs, \citet{santos2010} found positive correlations between the $S_{MW}$ activity index and BIS for early-K stars.
This is a curious result that should be investigated further.

\item 
We found that the long-term activity level was in general well correlated with FWHM. 43\% of the stars with long-term variability had their activity level significantly correlated with FWHM.
This is a good indication that the width of the CCF profile can be used to follow long-term activity in M dwarfs, as a complement to activity indices.
A similar result was found previously for the case of early-K dwarfs by \citet{santos2010}.

\item 
When compared with activity, the contrast showed a tendency for negative correlations with only 14\% of significant cases.
The tendency for anti-correlations was also found by \citet{santos2010} for early-K dwarfs, but since the fraction of significant coefficients is very low, this parameter does not seem to be a strong tool to diagnose long-term activity variability.

\end{itemize}

We found some cases of radial velocity variations induced by long-term activity. However, this does not seem to be a general trend.
This can be due to several reasons including the bad frequency sampling of observations (since this is a program dedicated to the search of planets, most of the observations were focused on short-term RV variability to find short-period planets or with bad high-frequency sampling when the aim was to detect long-period companions), short timespan of the data (although we found some activity cycle periods to be of the 3-year order some could have periods longer than the $\sim$6-year span of our observations), and the fact that some of the stars might probably have low-mass and/or long-period planets not yet detected (and thus their RV signal could be adding confusion to our activity study).
A longer timespan of observations will certainly contribute to a better understanding of how long-term activity cycles influence the detected velocity of these late-type stars.

In light of these results we must advise planet hunters to carefully check for long-term activity variations when faced with long-timespans of RV data since activity cycles could be adding noise or even mimicking the low-amplitude ($\leq$5 m\,s$^{-1}$) signals of planetary companions.

\acknowledgements{
This work has been supported by the European Research Council/European Community under the FP7 through a Starting Grant, as well as in the form of a grant reference PTDT/CTE-AST/098528/2008, funded by Funda\c{c}\~ao para a Ci\^encia e a Tecnologia (FCT), Portugal. J.G.S. would like to thank the financial support given by FCT in the form of a scholarship, namely SFRH/BD/64722/2009. N.C.S. would further like to thank the support from FCT through a Ci\^encia 2007 contract funded by FCT/MCTES (Portugal) and POPH/FSE (EC).}

\bibliographystyle{aa} 
\bibliography{magcyclesM_rv} 

\begin{thebibliography}{41}
\expandafter\ifx\csname natexlab\endcsname\relax\def\natexlab#1{#1}\fi

\bibitem[{{Bailey} {et~al.}(2009){Bailey}, {Butler}, {Tinney}, {Jones},
  {O'Toole}, {Carter}, \& {Marcy}}]{bailey2009}
{Bailey}, J., {Butler}, R.~P., {Tinney}, C.~G., {et~al.} 2009, \apj, 690, 743

\bibitem[{{Baliunas} {et~al.}(1995){Baliunas}, {Donahue}, {Soon}, {Horne},
  {Frazer}, {Woodard-Eklund}, {Bradford}, {Rao}, {Wilson}, {Zhang}, {Bennett},
  {Briggs}, {Carroll}, {Duncan}, {Figueroa}, {Lanning}, {Misch}, {Mueller},
  {Noyes}, {Poppe}, {Porter}, {Robinson}, {Russell}, {Shelton}, {Soyumer},
  {Vaughan}, \& {Whitney}}]{baliunas1995}
{Baliunas}, S.~L., {Donahue}, R.~A., {Soon}, W.~H., {et~al.} 1995, \apj, 438,
  269

\bibitem[{{Ballard} {et~al.}(2010){Ballard}, {Christiansen}, {Charbonneau},
  {Deming}, {Holman}, {Fabrycky}, {A'Hearn}, {Wellnitz}, {Barry}, {Kuchner},
  {Livengood}, {Hewagama}, {Sunshine}, {Hampton}, {Lisse}, {Seager}, \&
  {Veverka}}]{ballard2010}
{Ballard}, S., {Christiansen}, J.~L., {Charbonneau}, D., {et~al.} 2010, \apj,
  716, 1047

\bibitem[{{Boisse} {et~al.}(2011){Boisse}, {Bouchy}, {H{\'e}brard}, {Bonfils},
  {Santos}, \& {Vauclair}}]{boisse2011}
{Boisse}, I., {Bouchy}, F., {H{\'e}brard}, G., {et~al.} 2011, \aap, 528, A4+

\bibitem[{{Boisse} {et~al.}(2009){Boisse}, {Moutou}, {Vidal-Madjar}, {Bouchy},
  {Pont}, {H{\'e}brard}, {Bonfils}, {Croll}, {Delfosse}, {Desort}, {Forveille},
  {Lagrange}, {Loeillet}, {Lovis}, {Matthews}, {Mayor}, {Pepe}, {Perrier},
  {Queloz}, {Rowe}, {Santos}, {S{\'e}gransan}, \& {Udry}}]{boisse2009}
{Boisse}, I., {Moutou}, C., {Vidal-Madjar}, A., {et~al.} 2009, \aap, 495, 959

\bibitem[{{Bonfils} {et~al.}(2011){Bonfils}, {Delfosse}, {Udry}, {Forveille},
  {Mayor}, {Perrier}, {Bouchy}, {Gillon}, {Lovis}, {Pepe}, {Queloz}, {Santos},
  {S{\'e}grasan}, \& {Bertaux}}]{bonfils2011}
{Bonfils}, X., {Delfosse}, X., {Udry}, S., {et~al.} 2011, submitted

\bibitem[{{Bonfils} {et~al.}(2005){Bonfils}, {Forveille}, {Delfosse}, {Udry},
  {Mayor}, {Perrier}, {Bouchy}, {Pepe}, {Queloz}, \& {Bertaux}}]{bonfils2005}
{Bonfils}, X., {Forveille}, T., {Delfosse}, X., {et~al.} 2005, \aap, 443, L15

\bibitem[{{Bonfils} {et~al.}(2007){Bonfils}, {Mayor}, {Delfosse}, {Forveille},
  {Gillon}, {Perrier}, {Udry}, {Bouchy}, {Lovis}, {Pepe}, {Queloz}, {Santos},
  \& {Bertaux}}]{bonfils2007}
{Bonfils}, X., {Mayor}, M., {Delfosse}, X., {et~al.} 2007, \aap, 474, 293

\bibitem[{{Buccino} {et~al.}(2011){Buccino}, {D{\'{\i}}az}, {Luoni},
  {Abrevaya}, \& {Mauas}}]{buccino2011}
{Buccino}, A.~P., {D{\'{\i}}az}, R.~F., {Luoni}, M.~L., {Abrevaya}, X.~C., \&
  {Mauas}, P.~J.~D. 2011, \aj, 141, 34

\bibitem[{{Butler} {et~al.}(2004){Butler}, {Vogt}, {Marcy}, {Fischer},
  {Wright}, {Henry}, {Laughlin}, \& {Lissauer}}]{butler2004}
{Butler}, R.~P., {Vogt}, S.~S., {Marcy}, G.~W., {et~al.} 2004, \apj, 617, 580

\bibitem[{{Cincunegui} {et~al.}(2007){Cincunegui}, {D{\'{\i}}az}, \&
  {Mauas}}]{cincunegui2007a}
{Cincunegui}, C., {D{\'{\i}}az}, R.~F., \& {Mauas}, P.~J.~D. 2007, \aap, 461,
  1107

\bibitem[{{Dall} {et~al.}(2006){Dall}, {Santos}, {Arentoft}, {Bedding}, \&
  {Kjeldsen}}]{dall2006}
{Dall}, T.~H., {Santos}, N.~C., {Arentoft}, T., {Bedding}, T.~R., \&
  {Kjeldsen}, H. 2006, \aap, 454, 341

\bibitem[{{D{\'{\i}}az} {et~al.}(2007{\natexlab{a}}){D{\'{\i}}az},
  {Cincunegui}, \& {Mauas}}]{diaz2007a}
{D{\'{\i}}az}, R.~F., {Cincunegui}, C., \& {Mauas}, P.~J.~D.
  2007{\natexlab{a}}, \mnras, 378, 1007

\bibitem[{{D{\'{\i}}az} {et~al.}(2007{\natexlab{b}}){D{\'{\i}}az},
  {Gonz{\'a}lez}, {Cincunegui}, \& {Mauas}}]{diaz2007b}
{D{\'{\i}}az}, R.~F., {Gonz{\'a}lez}, J.~F., {Cincunegui}, C., \& {Mauas},
  P.~J.~D. 2007{\natexlab{b}}, \aap, 474, 345

\bibitem[{{Dumusque} {et~al.}(2011{\natexlab{a}}){Dumusque}, {Lovis},
  {S{\'e}gransan}, {Mayor}, {Udry}, {Benz}, {Bouchy}, {Lo Curto}, {Mordasini},
  {Pepe}, {Queloz}, {Santos}, \& {Naef}}]{dumusque2011a}
{Dumusque}, X., {Lovis}, C., {S{\'e}gransan}, D., {et~al.} 2011{\natexlab{a}},
  ArXiv e-prints

\bibitem[{{Dumusque} {et~al.}(2011{\natexlab{b}}){Dumusque}, {Udry}, {Lovis},
  {Santos}, \& {Monteiro}}]{dumusque2011b}
{Dumusque}, X., {Udry}, S., {Lovis}, C., {Santos}, N.~C., \& {Monteiro},
  M.~J.~P.~F.~G. 2011{\natexlab{b}}, \aap, 525, A140+

\bibitem[{{Endl} {et~al.}(2002){Endl}, {K{\"u}rster}, {Els}, {Hatzes},
  {Cochran}, {Dennerl}, \& {D{\"o}bereiner}}]{endl2002}
{Endl}, M., {K{\"u}rster}, M., {Els}, S., {et~al.} 2002, \aap, 392, 671

\bibitem[{{ESA}(1997)}]{esa1997}
{ESA}. 1997, VizieR Online Data Catalog, 1239, 0

\bibitem[{{Forveille} {et~al.}(2011{\natexlab{a}}){Forveille}, {Bonfils},
  {Delfosse}, {Alonso}, {Udry}, {Bouchy}, {Gillon}, {Lovis}, {Neves}, {Mayor},
  {Pepe}, {Queloz}, {Santos}, {Segransan}, {Almenara}, {Deeg}, \&
  {Rabus}}]{forveille2011a}
{Forveille}, T., {Bonfils}, X., {Delfosse}, X., {et~al.} 2011{\natexlab{a}},
  ArXiv e-prints

\bibitem[{{Forveille} {et~al.}(2009){Forveille}, {Bonfils}, {Delfosse},
  {Gillon}, {Udry}, {Bouchy}, {Lovis}, {Mayor}, {Pepe}, {Perrier}, {Queloz},
  {Santos}, \& {Bertaux}}]{forveille2009}
{Forveille}, T., {Bonfils}, X., {Delfosse}, X., {et~al.} 2009, \aap, 493, 645

\bibitem[{{Forveille} {et~al.}(2011{\natexlab{b}}){Forveille}, {Bonfils}, {Lo
  Curto}, {Delfosse}, {Udry}, {Bouchy}, {Lovis}, {Mayor}, {Moutou}, {Naef},
  {Pepe}, {Perrier}, {Queloz}, \& {Santos}}]{forveille2011b}
{Forveille}, T., {Bonfils}, X., {Lo Curto}, G., {et~al.} 2011{\natexlab{b}},
  \aap, 526, A141

\bibitem[{{Gomes da Silva} {et~al.}(2011){Gomes da Silva}, {Santos}, {Bonfils},
  {Delfosse}, {Forveille}, \& {Udry}}]{gomesdasilva2011}
{Gomes da Silva}, J., {Santos}, N.~C., {Bonfils}, X., {et~al.} 2011, \aap, 534,
  A30+

\bibitem[{{Hawley} {et~al.}(1996){Hawley}, {Gizis}, \& {Reid}}]{hawley1996}
{Hawley}, S.~L., {Gizis}, J.~E., \& {Reid}, I.~N. 1996, \aj, 112, 2799

\bibitem[{{Knutson} {et~al.}(2011){Knutson}, {Madhusudhan}, {Cowan},
  {Christiansen}, {Agol}, {Deming}, {D{\'e}sert}, {Charbonneau}, {Henry},
  {Homeier}, {Langton}, {Laughlin}, \& {Seager}}]{knutson2011}
{Knutson}, H.~A., {Madhusudhan}, N., {Cowan}, N.~B., {et~al.} 2011, \apj, 735,
  27

\bibitem[{{Koen} {et~al.}(2010){Koen}, {Kilkenny}, {van Wyk}, \&
  {Marang}}]{koen2010}
{Koen}, C., {Kilkenny}, D., {van Wyk}, F., \& {Marang}, F. 2010, \mnras, 403,
  1949

\bibitem[{{K{\"u}rster} {et~al.}(2003){K{\"u}rster}, {Endl}, {Rouesnel}, {Els},
  {Kaufer}, {Brillant}, {Hatzes}, {Saar}, \& {Cochran}}]{kurster2003}
{K{\"u}rster}, M., {Endl}, M., {Rouesnel}, F., {et~al.} 2003, \aap, 403, 1077

\bibitem[{{Lovis} {et~al.}(2011){Lovis}, {Dumusque}, {Santos}, {Bouchy},
  {Mayor}, {Pepe}, {Queloz}, {S{\'e}gransan}, \& {Udry}}]{lovis2011}
{Lovis}, C., {Dumusque}, X., {Santos}, N.~C., {et~al.} 2011, ArXiv e-prints

\bibitem[{{Mayor} {et~al.}(2009){Mayor}, {Bonfils}, {Forveille}, {Delfosse},
  {Udry}, {Bertaux}, {Beust}, {Bouchy}, {Lovis}, {Pepe}, {Perrier}, {Queloz},
  \& {Santos}}]{mayor2009}
{Mayor}, M., {Bonfils}, X., {Forveille}, T., {et~al.} 2009, \aap, 507, 487

\bibitem[{{Mayor} {et~al.}(2011){Mayor}, {Marmier}, {Lovis}, {Udry},
  {S{\'e}gransan}, {Pepe}, {Benz}, {Bertaux}, {Bouchy}, {Dumusque}, {Lo Curto},
  {Mordasini}, {Queloz}, \& {Santos}}]{mayor2011}
{Mayor}, M., {Marmier}, M., {Lovis}, C., {et~al.} 2011, ArXiv e-prints

\bibitem[{{Melo} {et~al.}(2007){Melo}, {Santos}, {Gieren}, {Pietrzynski},
  {Ruiz}, {Sousa}, {Bouchy}, {Lovis}, {Mayor}, {Pepe}, {Queloz}, {da Silva}, \&
  {Udry}}]{melo2007}
{Melo}, C., {Santos}, N.~C., {Gieren}, W., {et~al.} 2007, \aap, 467, 721

\bibitem[{{Meunier} {et~al.}(2010){Meunier}, {Desort}, \&
  {Lagrange}}]{meunier2010}
{Meunier}, N., {Desort}, M., \& {Lagrange}, A.-M. 2010, \aap, 512, A39+

\bibitem[{{Moutou} {et~al.}(2011){Moutou}, {Mayor}, {Lo Curto},
  {S{\'e}gransan}, {Udry}, {Bouchy}, {Benz}, {Lovis}, {Naef}, {Pepe}, {Queloz},
  {Santos}, \& {Sousa}}]{moutou2011}
{Moutou}, C., {Mayor}, M., {Lo Curto}, G., {et~al.} 2011, \aap, 527, A63+

\bibitem[{{Queloz} {et~al.}(2009){Queloz}, {Bouchy}, {Moutou}, {Hatzes},
  {H{\'e}brard}, {Alonso}, {Auvergne}, {Baglin}, {Barbieri}, {Barge}, {Benz},
  {Bord{\'e}}, {Deeg}, {Deleuil}, {Dvorak}, {Erikson}, {Ferraz Mello},
  {Fridlund}, {Gandolfi}, {Gillon}, {Guenther}, {Guillot}, {Jorda}, {Hartmann},
  {Lammer}, {L{\'e}ger}, {Llebaria}, {Lovis}, {Magain}, {Mayor}, {Mazeh},
  {Ollivier}, {P{\"a}tzold}, {Pepe}, {Rauer}, {Rouan}, {Schneider},
  {Segransan}, {Udry}, \& {Wuchterl}}]{queloz2009}
{Queloz}, D., {Bouchy}, F., {Moutou}, C., {et~al.} 2009, \aap, 506, 303

\bibitem[{{Queloz} {et~al.}(2001){Queloz}, {Henry}, {Sivan}, {Baliunas},
  {Beuzit}, {Donahue}, {Mayor}, {Naef}, {Perrier}, \& {Udry}}]{queloz2001}
{Queloz}, D., {Henry}, G.~W., {Sivan}, J.~P., {et~al.} 2001, \aap, 379, 279

\bibitem[{{Saar} \& {Donahue}(1997)}]{saar1997a}
{Saar}, S.~H. \& {Donahue}, R.~A. 1997, \apj, 485, 319

\bibitem[{{Saar} \& {Fischer}(2000)}]{saar2000}
{Saar}, S.~H. \& {Fischer}, D. 2000, \apjl, 534, L105

\bibitem[{{Santos} {et~al.}(2010){Santos}, {Gomes da Silva}, {Lovis}, \&
  {Melo}}]{santos2010}
{Santos}, N.~C., {Gomes da Silva}, J., {Lovis}, C., \& {Melo}, C. 2010, \aap,
  511, A54+

\bibitem[{{Santos} {et~al.}(2000){Santos}, {Mayor}, {Naef}, {Pepe}, {Queloz},
  {Udry}, \& {Blecha}}]{santos2000}
{Santos}, N.~C., {Mayor}, M., {Naef}, D., {et~al.} 2000, \aap, 361, 265

\bibitem[{{S{\'e}gransan} {et~al.}(2011){S{\'e}gransan}, {Mayor}, {Udry},
  {Lovis}, {Benz}, {Bouchy}, {Lo Curto}, {Mordasini}, {Moutou}, {Naef}, {Pepe},
  {Queloz}, \& {Santos}}]{segransan2011}
{S{\'e}gransan}, D., {Mayor}, M., {Udry}, S., {et~al.} 2011, ArXiv e-prints

\bibitem[{{Udry} {et~al.}(2007){Udry}, {Bonfils}, {Delfosse}, {Forveille},
  {Mayor}, {Perrier}, {Bouchy}, {Lovis}, {Pepe}, {Queloz}, \&
  {Bertaux}}]{udry2007}
{Udry}, S., {Bonfils}, X., {Delfosse}, X., {et~al.} 2007, \aap, 469, L43

\bibitem[{{Zechmeister} {et~al.}(2009){Zechmeister}, {K{\"u}rster}, \&
  {Endl}}]{zechmeister2009}
{Zechmeister}, M., {K{\"u}rster}, M., \& {Endl}, M. 2009, \aap, 505, 859

\end{thebibliography}

\listofobjects

\end{document}